\documentclass[aoas,preprint]{imsart}

\RequirePackage[OT1]{fontenc}
\RequirePackage{amsthm,amsmath}

\RequirePackage{natbib}
\RequirePackage[colorlinks,citecolor=blue,urlcolor=blue]{hyperref}

\usepackage{array}
\usepackage{longtable}
\usepackage{capt-of}
\usepackage{float}
\usepackage{amssymb}
\usepackage{graphicx}
\usepackage{upgreek}
\usepackage{tikz}

\usepackage{prodint}
\usepackage{mathtools}
\usepackage{dsfont}  
\usepackage{calligra} 
\usepackage{stackengine}

\usepackage{algorithm}
\usepackage[noend]{algpseudocode}

\startlocaldefs

\theoremstyle{plain}
\newtheorem{thm}{Theorem}[section]

\theoremstyle{plain}

\theoremstyle{plain}
\newtheorem{prop}[thm]{Proposition}

\theoremstyle{definition}
\newtheorem{defn}[thm]{Definition}

\theoremstyle{remark}
\newtheorem*{rem*}{Remark}

\endlocaldefs

\begin{document}

\begin{frontmatter}
\title{Efficiency in Lung Transplant Allocation Strategies}
\runtitle{Lung Transplant Allocation Strategies}

\begin{aug}
\author{\fnms{Jingjing} \snm{Zou}\thanksref{m1}\ead[label=e1]{j2zou@health.ucsd.edu}},
\author{\fnms{David J.} \snm{Lederer}\thanksref{m2}\ead[label=e2]{dl427@cumc.columbia.edu}}
\and
\author{\fnms{Daniel} \snm{Rabinowitz}\thanksref{m3}\ead[label=e3]{dan@stat.columbia.edu}}
\runauthor{Zou et al.}

\affiliation{University of California, San Diego\thanksmark{m1}, Regeneron Pharmaceuticals, Inc.\thanksmark{m2}, Columbia University Irving Medical Center\thanksmark{m2} and Columbia University\thanksmark{m3}}

\address{Address of the First author\\
	Division of Biostatistics and Bioinformatics \\
	Department of Family Medicine and Public Health \\
	University of California, San Diego \\
	La Jolla, CA, USA 92093 \\
\printead{e1}}

\address{Address of the Second author\\
(1). 777 Old Saw Mill River Rd \\
Tarrytown, NY, USA 10591 \\
(2). 622 W 168th St, PH-8, Room 101\\
New York, NY, USA 10032\\
\printead{e2}}

\address{Address of the Third author\\
	1025 Amsterdam Ave. 10th Floor\\
	New York, NY, USA, 10027\\
\printead{e3}}
\end{aug}

\begin{abstract}
Currently in the United States, lung transplantations are allocated to candidates according to the candidates' Lung Allocation Score (LAS).
The LAS is an ad-hoc ranking system for patients' priorities of transplantation.
The goal of this study is to develop a framework for improving patients' life expectancy over the LAS based on a comprehensive modeling of the lung transplantation waiting list.
Patients and organs are modeled as arriving according to Poisson processes, patients’ health status evolving  a waiting time inhomogeneous Markov process until death or transplantation, with organ recipient's expected post-transplant residual life depending on waiting time and health status at transplantation.
Under allocation rules satisfying minimal fairness requirements, the long-term average expected life converges, and its limit is a natural standard for comparing allocation strategies.
Via the Hamilton-Jacobi-Bellman equations, upper bounds for the limiting average expected life are derived as a function of organ availability. Corresponding to each upper bound is an allocable set of (state, time) pairs at which patients would be optimally transplanted. The allocable set expands monotonically as organ availability increases, which motivates the development of an allocation strategy that leads to long-term expected life close to the upper bound.
Simulation studies are conducted with model parameters estimated from national lung transplantation data. Results suggest that compared to the LAS, the proposed allocation strategy could provide a 7.7\% increase in average total life.
We further extended the results to the case of multiple-organ-type matching.

\end{abstract}


\end{frontmatter}

\section{Introduction}

\subsection{The Lung Transplantation Score (LAS)}
Lung transplantation was first performed in 1963 and has evolved over the years to become a standard treatment for patients with advanced lung diseases (\cite{Kotloff:2012jf}). 
There are not enough organs available for transplantation to meet the need of patients waiting for lungs, and efficient allocation of lung transplantations to patients is vital. 

Currently in the United States, patients in need of lung transplantation are registered to waiting lists managed by the United Network for Organ Sharing (UNOS). As lungs become available for transplant, they are allocated to candidates in waiting lists on the basis of age, geography, blood type (ABO) compatibility, and the Lung Allocation Score (LAS). 
The LAS was first implemented in 2005, aiming to reduce the risk of in-waiting-list mortality and to prolong post-transplant residual life of organ recipients (\cite{Valapour:2017cr}). 
The LAS is calculated as 
$100\cdot(\textrm{PTAUC}-2\cdot\textrm{WLAUC}+730)/1095$,
where WLAUC is the estimated in-waiting-list life expectancy during an additional year if not transplanted and PTAUC is the estimated post-transplant life expectancy during the first year.
Both in-waiting-list and post-transplant survivals are computed using proportional hazards models given current values of relevant characteristics of patients. 
The expected numbers of days of survival are calculated by integrating the areas under the covariate-specific survival curves within the first year, under the assumption that a patient's health status would remain constant. 
When organs become available, patients in the waiting list are ranked according to their LAS values. Candidates with higher LAS values are given higher priorities for transplantation (\cite{UNOS:2015wn}).

The LAS is recognized to be imperfect. 
In February 2015, the Organ Procurement and Transplantation Network (OPTN) managed by UNOS implemented a revision of the LAS. Modifications were made to the covariates and their weights in the calculation to better predict patients' survival (\cite{Valapour:2017cr}).

There are other concerns about the LAS besides the selection of covariates.
Studies have demonstrated that the emphasis on reducing the risk of in-waiting-list mortality in the LAS calculation may have the effect of increasing post-transplant mortality, since patients with the highest LAS are often in the worst conditions and subject to high post-transplant risk of death (\cite{Liu:2010gr}, \cite{Russo:2010kb,Russo:2011bk}, \cite{Merlo:2009dm}). 
And hence the question of whether the patients prioritized by the LAS are those who would benefit the most from transplantation.
Studies also suggested that the LAS focused on the 1-year survival and failed to assess the long-term benefit of lung transplantation for the recipients. Results in \cite{Maxwell:2014kf} showed that the 5-year survival had in fact slightly decreased after the implementation of the LAS.

\subsection{Previous Studies on Organ Transplantation}

Organ allocation has been studied from the policy maker's perspective.
The goal is generally characterized as assignments of organs to patients in the waiting list that optimize expected outcomes. Early works can be traced back to \cite{Derman:1971ey, Derman:1975vd}, \cite{Albright:1972ck} and \citet{RIGHTER:1989ds}, in which the authors solved for the optimal solution for allocating sequentially arrived resources to a finite number of subjects. The focus of these studies was on general resource allocation problems and for this purpose, the modeling was simplified and not tailored specifically to the organ transplantation problem. 

Recent studies have involved increasing model complexity in studying the organ allocation problem.
\cite{Zenios:1999uj} and \cite{Zenios:1999ba,Zenios:2000dl} modeled the waiting list for kidney transplantation with a deterministic fluid model: patients and organ donors were categorized into classes based on their demographic, immunological, and physiological characteristics and different categories of patients and organs flow in and out of the waiting list at class-specific rates.
An optimal allocation rule was derived under the standard of a linear combination of quality-adjusted life expectancy and equity.
\cite{Akan:2012cj} modeled the waiting list for donated livers with a fluid model, in which patients were categorized into multiple classes according to their health status. 
Patients in one class were assumed to be able to flow only into adjacent categories with deterministic rates.
Optimal allocation strategies were developed under a standard combining expected total number of in-waiting-list deaths and quality-adjusted total life expectancy. 

Organ allocation had also been studied from an individual patient's perspective by, for example, \cite{Ahn:1996it}, \cite{Hornberger:1997ez}, \cite{Su:2004jn,Su:2005eu,Su:2006uq}, \cite{Alagoz:2004dd, Alagoz:2007cl,Alagoz:2007iv}), and \cite{Ata:2018jg}. These studies focused on optimal patient strategies for accepting or rejecting offers of organs.

Aspects of the organ transplantation problem other than organ allocation have also been studied. For example, the impact of available cadaveric kidneys on the number of candidates in the waiting list (\cite{RUTH:1985wj}), kidney exchange programs (\citet{Roth:2004ue,Roth:2005bi}, \citet{Ashlagi:2011gk,Ashlagi:2012wq}, \citet{Cechlarova:2012il}), the design of policies that meet fairness constraints chosen by the policy maker (\citet{Bertsimas:2013ey}), and comparing the efficiency and fairness of three proposed heart allocation policies via simulation studies (\cite{Hasankhani:2017da}). 
In one of the few studies on lung transplantation, \citet{Anonymous:08ugtNlN} evaluated the survival benefit of lung transplantation with the LAS from a causal inference perspective.

\subsection{Overview of Model and Results}

Relative to the richness of the literature on kidney and liver transplantation, few attempts have been made to study organ allocation in lung transplantation. 
Yet models for kidney and liver transplantation waiting lists cannot be appropriated wholesale in the service of modeling lung transplantation as several characteristics distinguish the transplantation of lungs from the transplantation of other organs.
Here we propose a model for the lung transplantation waiting list based on common characteristics shared by different organ types and on the characteristics unique to lungs.

A waiting list for lung transplantation consists of patients added at random calendar times with different characteristics such as diagnosis, demographics, and health indicators. Here patient arrivals are modeled by a calendar time homogeneous Poisson process. Patients' initial states upon arrival are modeled as sampled randomly from a finite state space.

Patients' states, especially their health status, may change during their sojourn on the waiting list.
As patients' states change, their hazards for in-waiting-list death and expected post-transplant residual life change accordingly, which in turn affect patients' priorities for transplantation.
Given the significant implications of changes in patients' states on allocation decisions, it is crucial to model the trajectory of patients' states accurately and comprehensively.

Here we model counterfactual patient health status trajectories, that is, trajectories would be observed without transplantations, as independent and identically distributed continuous-time Markov processes.
The model proposed here reflects UNOS practice and that patients evolve in continuous time, and differs from those in the previous studies, in that a patient's health status trajectory is indexed by the waiting time since the patient arrives, and the change in health status is assumed to be inhomogeneous with respect to waiting time. 
Waiting time since listing is an important indicator for a patient's functional age, which is increasingly accepted in predicting potential outcomes (\citet{Kotloff:2012jf}). 
The inhomogeneous assumption enables the model to capture changes in the health status transition rate as waiting time increases.
A scenario that would benefit from the proposed modeling, for instance, would be when patients experience accelerated deterioration in health after waiting for a long period in a serious condition.
In addition, instead of only allowing transitions between adjacent states, the proposed model allows the consideration of more complex patterns of transitions in patients' health status.

Organs, either from deceased or living donors, also become available at random calendar times. 
Here organ arrivals are modeled as a homogeneous Poisson process that is independent of patient arrivals and transitions in health status. 
When an organ becomes available, a patient in the waiting list is selected for transplantation.
Due to the short time between when an organ becomes available and when the organ is no longer viable for transplantation, available organs are transplanted almost instantly to the selected patient.
An organ recipient's health status at the time of transplantation affects the post-transplantation residual life (\citet{UNOS:2011tt}). 
Candidates who are not transplanted will remain in the waiting list for future transplantation opportunities. Except in rare circumstances, patients leave the waiting list only in case of transplantation or death.

In contrast to the case of kidney and liver transplantation, living donor transplantation is extremely rare in lung transplantation: according to \citet{UNOS:2011tt}, 8,674 patients were in the waiting list from year 2009 to 2011, 5,172 received deceased donor transplant and only two received living donor transplant.
Therefore our focus here is on deceased donor transplantation only. 
Statistics in \citet{UNOS:2011tt} also show that from 2009 to 2011 only 20 patients out of 5,192 refused the transplantation offer, indicating patient choice is practically negligible in modeling lung transplantation. Therefore, we ignore the possibility of organ refusal and focus on the policy maker's perspective in studying lung allocation.
 
The policy maker uses allocation rules to determine which patient in the current waiting list is selected for transplantation when an organ is available. 
Allocation rules may make use of any information of the current state or history of the patients in the waiting list. 
There are ethical issues to be considered regarding fairness or equity. \cite{Zenios:2000dl} defined absolute equity with which the discrepancies in the outcome across patient groups would be eliminated and relative equity in which the first-come-first-transplanted policy was used as the golden standard.
\cite{Su:2006uq} and \cite{Ata:2018jg} defined equity as equal chances of receiving an organ for all patient groups.
\cite{Akan:2012cj} suggested to remove factors such as age, gender and race in making allocation decisions.

In our study, allocation rules that are not consistent over time or that are affected by factors other than waiting times and relevant health characteristics of patients are viewed as inequitable. 
Relevant health characteristics are those that predict outcomes by which allocation rules are compared, such as the current and future hazards for death and/or post-transplantation survival.
Fair allocations should depend on waiting times and health characteristics of the patients in the waiting list at the time of organ arrivals and possible independent randomizations. 
Allocation rules obtained by defining an index as a function of a patient's waiting time and relevant characteristics and allocating to the patient in the list with the highest index value (or randomly among patients tied for the highest value) are guaranteed to be fair.

Some standards for comparing allocation rules that have been considered may be characterized in terms of expectations. 
Defining a standard in terms of an expectation, however, does not unambiguously define an optimization problem. Such a definition begs the question of which patients' expectations are to be maximized. Implicitly, such standards refer to the long-term averages of life quantities of patients in the waiting list as time goes to infinity. 

Here we take the expected total life (in-waiting-list life plus post-transplant life) averaged over all patients as calendar time increases as the metric for evaluating allocation strategies. 
We show that under the minimal fairness constraints, the waiting list has a unique limiting distribution.
On average, patients entering the waiting list are transplanted according to a unique allocation-rule-specific limiting transplantation rate, which is a function of waiting time and health status. 
The long-term average of patients' total life exists and is a functional of the limiting transplantation rate. 
The method used here can be extended readily to include the long-term average of essentially any aspect of patient expectation.

It is also shown that the expected proportion of transplanted patients is bounded by the ratio of the intensity of organ arrivals to that of patient arrivals, and that the transplantation rate satisfies boundedness constraints related to the counterfactual transition rates.
We begin the search for the optimal fair allocation rule by solving for the limiting waiting time and health status specific transplantation rate that optimizes the long-term average life subject to the boundedness constraints.

The Hamilton-Jacobi-Bellman equations are used to characterize the form of the optimal limiting transplantation rate.  
The optimal rate is nonzero whenever the difference between the expected residual life with an immediate transplant and the expected residual life without an immediate transplant is greater than one minus the probability of future transplantations, scaled by a penalty parameter associated with the ratio of organ arrival rate to patient arrival rate. 

Not every transplantation rate satisfying the constraints is a limiting rate for some allocation strategy. In particular, the rate corresponding to the optimal solution may not be achievable, as patients in the waiting list that are alive and not transplanted do not necessarily include those at waiting times and states where the optimal rate is non-zero when an organ arrives. 

Here we propose an allocation strategy that is designed so that the corresponding waiting time and health status specific transplantation rate is close to the rate given by the optimal solution. The proposed allocation strategy relies on a critical monotonicity property: as the penalty parameter associated with organ availability decreases, combinations of waiting time and health state become allocable with non-zero optimal transplantation rate in a monotonic manner, and the order of each combination becoming allocable makes a natural index of transplantation priority.

A comprehensive simulation study is conducted with model parameters estimated from the national lung transplantation data provided by UNOS to examine the effect of the proposed strategy comparing to the current LAS system. 
Results suggest that the proposed allocation strategy can provide a gain of at least 7\% in expected average total life relative to the LAS.

We further extend the single-organ-type results to the case of multiple organ types. The optimal transplantation rates for all organ types are derived and an allocation strategy to match and allocate organs of each type to patients is developed based on monotonicity properties of the optimal allocation rates.

We would also like to note that although our focus in this article is on lung transplantation, the model and methods developed can be applied to a wide range of applications, including the transplantation and matching of other organ types and allocations of scarce (medical) resources to candidates in complex transitions of states in general.

In what follows,
Section \ref{sect: model} introduces notation in modeling the lung transplantation waiting list.
Section \ref{sect: optimization} characterizes the comparison of allocation rules in terms of a constrained optimization problem and states the optimal solution. 
Section \ref{sect: strategy} proposes a practical allocation strategy in terms of an allocation index based on the optimal solution.
Section \ref{sect: multi} extends the results to the multiple-organ-type case.
Section \ref{sect: comparison} compares average total life of patients with the proposed allocation strategy and the LAS in simulation studies.
Section \ref{sec: multiple numeric} provides further numerical results of the multiple organ type scenario.
Proofs of the theoretical results are postponed to the Appendix and Supplementary Material.

\section{The Waiting List} \label{sect: model}

Let $\tau<\infty$ denote the intensity of the Poisson process of patient arrivals to the waiting list, and let $0<T_{1}<T_{2}<\dots$ denote the patients' arrival times. For convenience, let $ N_{t} $ denote the number of patient arrivals up to calender time $ t $. 
Let $\rho<\infty$ denote the intensity of the Poisson process of organ arrivals, and let $0<S_{1}<S_{2}<\dots$ denote the arrival times of organs. Let $ O_{t} $ denote the number of organ arrivals up to calendar time $ t $. Assume organ arrivals are independent of patient arrivals. We are interested in settings when $\rho<\tau$, as otherwise the supply of organs would generally meet the demand, and there would be no need for an allocation rule.

Let $\mathcal{X}=\{0,1,...,n\}$ denote the finite set of possible patient health status, in which $0$ denotes the absorbing state corresponding to death.
Let $\{X(s):s \geq 0 \}$ denote a generic health status trajectory indexed by waiting time since arrival. 
Suppose $ X(s) $ is sufficiently detailed that $\{X(s):s\geq 0\}$ is a c\`adl\`ag Markov process.
Denote the transition kernel of the process by $p_{s,t}(i,j)$, and the infinitesimal generator by $q_{ij}(s)=\lim_{t\downarrow s}(p_{s,t}(i,j)-I{(i=j)} )/(t-s)$, where $ I{(\cdot )}$ denotes the indicator function. 
Relative to the granularity of measurements of patient health status, there are no patient states that lead inevitably to sudden transitions, so it may safely be assumed that the max total transition rate, $\sum_{j\neq i}q_{ij}(s)<\infty$. 
In addition the expected number of transitions
in any finite waiting time interval is finite, for which a sufficient condition might be $ \int_0^T q_{ij}(s)ds <\infty$ for any $ i,j \in \mathcal{X} $. 
Let the vector $\boldsymbol{p} = (p_{1},\dots,p_{n})$
denote the distribution of patients' initial states upon arrival, where $p_{i}=P(X(0)=i)$ for each $i\in\mathcal{X}$. As patients must enter the waiting list alive, $\sum_{i=1}^n p_i = 1$.
Let $T$ be the upper bound of a patient's waiting time, i.e., $P(X(s)=0) = 1$ for all $s\geq T$. 

The effect of allocation to a patient is measured by the difference between the patient’s expected residual life with a transplant and expected residual life without a transplant.
Let $R(s)$ denote a generic post-transplant residual life if transplantation occurs at waiting time $ s \geq 0$.
Assume that the characterization of patient health states is sufficiently informative such that for any $ s \ge 0 $, $ R(s) $ and $ \sigma(\{X(u):u < s \} ) $ are conditionally independent given $ X(s) $. Since post-transplant residual life is non-negative and bounded, $ R(s) \ge 0$ and $\sup_{s \ge 0,i\in \mathcal{X}} E\left( R(s) \mid X(s) = i \right) <\infty$.

Let $\{(X^{(k)}(s),R^{(k)}(s)): s \ge 0 \}$ denote the counterfactual health status trajectories and post-transplant residual life processes of the $ k $th arriving patient. $\{(X^{(k)}(s),R^{(k)}(s)): s \ge 0,\, k\in \mathbb{N} \}$ are independent and identically distributed copies of the generic pair $\{(X(s), R(s)): s \ge 0\}$, and are independent of $ \{T_i, S_j: i,j\in \mathbb{N} \} $. 
  
An allocation sequence $\{a_j: j = 1, 2, \dots \}$ is a random sequence of patient indices, where $ a_j $ is the index of the recipient of the $ j $th organ. Organs are allocated promptly after being retrieved to keep their functionality, and the short delay in allocation after organ arrival is omitted. For each $j$, $T_{a_{j}} \le S_{j}$, $X^{(a_{j})}(S_{j}-T_{a_{j}})\neq 0$, and $a_{j}\neq a_{k}$ if $j\neq k$, which reflects that patients who receive transplantation must have entered the waiting list before the organ and must be alive at the moment of allocation, and that patients exit the waiting list upon receiving a transplant.

Let $ T_{T}^{(k)} $ denote the $ k $th patient's waiting time at transplantation, that is, $ T_{T}^{(k)} = S_j - T_k $ if $ a_j = k $ and $ T_{T}^{(k)} = \infty $ if $ k \notin \{a_j: j\in \mathbb{N} \} $.
Let $(n+1)$ denote the post-transplant state and let $\mathcal{X}_{0}=\mathcal{X} \cup \{n+1\}$ denote the augmented patient state space. 
Denote the $k$th patient's actual trajectory by $ \{\tilde{X}^{(k)}(s): s\ge 0 \} $, so that
\begin{equation} 
\tilde{X}^{(k)}(s) = X^{(k)}(s)\cdot I{ (s<T_{T}^{(k)})} + (n+1) \cdot I{ (s \ge T_{T}^{(k)} )}, 
\end{equation}
and $\tilde{X}^{(k)}(s)$ is a c\`adl\`ag process on $ \mathcal{X}_{0} $.

To formulate the waiting list, we first define a filtration that describes the information relevant to events in the waiting list up to time $ t $, including patient and organ arrivals, patients' counterfactual health state transitions and allocation decisions: 
\begin{equation}
\mathcal{F}_{t}=\sigma \big(T_{k},S_{i},a_{i},\{X^{(k)}(s):s\in[0,t-T_{k}]\}:T_{k},S_{i}\in[0,t] \big).
\end{equation}
At any time, the current waiting list consists of waiting time and health status of patients who have arrived and have not died nor transplanted. 
Let $w_{0}$ denote the state of no patient in the list. 
On $\mathcal{W}=\cup_{d=1}^{\infty}([0,T]\times\mathcal{X})^{d} \cup \{w_0\}$ and its Borel $\sigma$-algebra $\mathcal{B}(\mathcal{W})$, define the waiting list process
\begin{equation}
W_{t}=\{(t-T_{k},\tilde{X}^{(k)}(t-T_{k})):T_{k}\leq t,\tilde{X}^{(k)}(t-T_{k})\notin\{0,n+1\} \}. \label{eq: defn WL}
\end{equation}
$ \{W_t: t\ge 0\} $ is adapted to $\{\mathcal{F}_{t}\}$. Since discontinuities of $\{W_{t}:t\geq0\}$ can only be a result of patient or organ arrivals or transitions in health status, which are all c\`adl\`ag, $W_{t}$ is c\`adl\`ag. 

We define another filtration generated by events in the waiting list up to time $ t $,  but excluding events associated with the allocation decision at $ t $.
\begin{equation}
\mathcal{G}_{t}=\sigma \big(T_{k},S_{i},\{X^{(k)}(s):s\in[0,t-T_{k}]\}:T_{k},S_{i}\in[0,t] \big)  
\vee \sigma(a_i: S_i<t),
\end{equation}
Denote the waiting list before potential allocation at calendar time $ t $ by 
\begin{multline}
W'_{t}=\{(t-T_{k}, X^{(k)}(t-T_{k})): T_k \le t, \, k \notin \{a_i:S_i < t\},\, X^{(k)}(t-T_{k}) \ne 0 \}.\label{eq: defn WL prime}
\end{multline}
For any $j\in\mathbb{N}$, $S_{j}$ is a $\{\mathcal{G}_{t}\}$-stopping time. Let $ \mathcal{G}_{S_{j}}=\{A:A\cap\{S_{j}\leq t\}\in\mathcal{G}_{t},\,\forall\, t\geq0\} $, then $ W'_{S_{j}} $ is measurable with respect to $ \mathcal{G}_{S_{j}} $ as $ W'_{s} $ is progressively measurable.

Allocation sequences should satisfy fairness requirements.
Our definition of fairness requires that given the waiting list up to the moment of allocation, the choice of the organ recipient should only depend on waiting time and health status of patients who are currently alive and not transplanted, and a randomization that is conditionally independent, given the current state of the waiting list, of the history of the waiting list. In formal terms,

\begin{defn} 
An allocation sequence $\{a_{1},a_{2},\dots\}$ is termed fair, if there exists a function $ \Gamma(\cdot, \cdot) $, such that 
\begin{enumerate}
\item for any $ x \in [0,T]\times\mathcal{X} $, $\Gamma(x, \cdot)$ is a measurable function on $(\mathcal{W},\mathcal{B}(\mathcal{W}) )$, and for any $ w \in \mathcal{W} \backslash \{w_0\} $, $\Gamma(\cdot, w)$ a probability measure on $ \{x: x \in w \} $,
\item for any $j\in\mathbb{N}$ and $A\in\mathcal{B}(\mathcal{W})$,
\begin{equation} \label{eq: fairness}
P((S_j-T_{a_j}, X^{(a_j)} (S_j-T_{a_j})) = x \mid \mathcal{G}_{S_{j}})=\Gamma(x, W'_{S_{j}}). 
\end{equation}  
\end{enumerate}
\end{defn} 

In the above definition, $ \Gamma $ is the allocation rule that determines the probabilities of allocating an available organ to patients in the current waiting list. Note the definition implies that given the waiting times and health status of patients in the current waiting list, the allocation probabilities are invariant to patient indices and consistent with respect to the calendar time.

Moreover, only realistic fair allocation rules that are non-informative of patients' post-transplant residual life and future events in the waiting list are considered, and thus 
\begin{multline} \label{eq: fairness2} 
P((S_j-T_{a_j}, X^{(a_j)} (S_j-T_{a_j})) = x \\ \mid W'_{S_{j}}, \{S_i, T_k: i,k \in \mathbb{N} \}, \{X^{(k)}(s), R^{(k)}(s):s \ge 0, k\in \mathbb{N} \}) \\
= \Gamma(x, W'_{S_{j}}).
\end{multline}

The effect of an allocation rule $ \Gamma $ is measured by the long-term average expected total life, including life in waiting list and post-transplantation, of all patients ever enter the waiting list.
The $k$th patient's total life is equal to 
\begin{equation} \label{eq: total_life}
\inf_{t\ge 0} \{X^{(k)}(t)=0 \} \wedge T^{(k)}_T + \sum_{i\in \mathcal{X}} \int_{0}^{T}R_{i}^{(k)}(s)d\tilde{N}_{i,n+1}^{(k)}(s), 
\end{equation}
where $ d\tilde{N}_{i,n+1}^{(k)}(s) = 1 $ if $T^{(k)}_T =s$ and $ d\tilde{N}_{i,n+1}^{(k)}(s) = 0 $ otherwise. 

It is shown in \cite{Zou:2015ur} that \eqref{eq: total_life} can be rewritten as
\begin{equation} \label{eq: total_life_rewrite}
\sum_{i=1}^n\int_{0}^{T}\tilde{R}^{(k)}(s)d\tilde{N}_{i,n+1}^{(k)}(s) + \inf_{t\ge 0} \{X^{(k)}(t)=0 \},
\end{equation}
where $ \tilde{R}^{(k)}(s) = R^{(k)}(s)- \big(\inf_{t\ge 0} \{X^{(k)}(t)=0\} - s \big) $
denotes the difference between the $k$th patient's residual life with transplantation at time $s$ and residual life if never transplanted. Note that the last term in \eqref{eq: total_life_rewrite} is invariant to allocation rules, and therefore the expected long-term average total life is equal to the long-term average of expected life gain from transplantation 
\begin{equation} \label{eq: life gain}
\lim_{t\rightarrow \infty} \frac{1}{N_t} E \Bigg[\sum_{j=1}^{O_t} \tilde{R}^{(a_j)}(S_j - T_{a_j}) \Bigg],
\end{equation}
plus a constant that is invariant to allocation rules. For simplicity of notation, we will use \eqref{eq: life gain} as the objective function in searching for optimal allocation rules.

\section{Optimizing the Average Expected Life Gain} \label{sect: optimization}

\subsection{Formulation of the Optimization Problem}

In studying the effect of allocation rule $ \Gamma $  on the long-term average expected life gain, a pivotal quantity is the long-term average occupancy of health state $i$ for $i \in \mathcal{X}$:
\begin{equation}  \label{eq: pi}
\pi_i(s) 
= \lim_{t\rightarrow \infty} \frac{1}{N_t} \sum_{k=1}^{N_t} I{ (\tilde{X}^{(k)}(s) = i)}.
\end{equation}
The long-term average of expected life gain can be expressed as a function of the long-term occupancy, and the occupancy can be shown to satisfy certain constraints - so that an upper bound for the long-term expected life gain may be found by maximizing over the long-term occupancy, subject to the constraints.

A proof is outlined in \cite{Zou:2015ur} for the proposition that, with any fair allocation rule, the limit in \eqref{eq: pi} exists so that $ \pi_i(s) $ is well defined for all $ i \in \mathcal{X} $ and $ s \in [0, T] $. The existence of the limit follows from the fact that with a fair allocation rule, the waiting list process $ \{W_t: t\ge 0\} $ is strong Markov with respect to filtration $ \{\mathcal{F}_t\} $ and is positive recurrent. Therefore by the ergodic theory there exists a finite invariant measure $ \sigma $ and $ \pi_i(s) $ can be expresses in terms of $ \sigma $.

For any fair allocation rule, there exists a corresponding long-term transplantation rate
\begin{equation} \label{eq: defn of Xi}
d\boldsymbol{\Psi}_{s}=\boldsymbol{\uppi}_{s-}\mathbf{Q}_{s}ds-d\boldsymbol{\uppi}_{s},
\end{equation}
where $ \mathbf{Q} = \{q_{ij} \} $ is the matrix of counterfactual rate of transition in health status.
The transplantation rate is the difference between the rate of evolution that would occur absent organ allocation and the rate of the evolution of occupancy with allocation. For $ s = 0 $, let $ \pi_i(s-) = P(X(0) = i) $. For achievable long-term occupancies, $ \pi_i(s) $ is smooth. The definition may be extended to the non-smooth case, which is relevant at the upper bound, by taking $ \Psi_i(\{s\}) = \pi_i(s-) - \pi_i(s) $ and allowing $ \Psi_i(\{s\}) > 0 $.

The effect of an allocation rule on the long-term average of expected life gain \eqref{eq: life gain} can be expressed in terms of the long-term transplantation rate. 
Let $\tilde{\mu}_{i}(s)=E(\tilde{R}(s) \mid X(s) = i)$ and denote $\tilde{\boldsymbol{\upmu}}_s = (\tilde{\mu}_{1}(s),\dots, \tilde{\mu}_{n}(s))$.
The result is stated formally in the following theorem. 
\begin{thm} \label{thm: expected life gain}
\normalfont
\begin{equation} \label{eq: expected life gain}
\lim_{t\rightarrow \infty} \frac{1}{N_t} {E} \Bigg[\sum_{j=1}^{O_t} \tilde{R}^{(a_j)}(S_j - T_{a_j}) \Bigg] = \int_0^T d\boldsymbol{\Psi}_{s} \tilde{\boldsymbol{\upmu}}_s.
\end{equation}
\end{thm}

While each fair allocation rule has a corresponding transplantation rate $ \boldsymbol{\Psi} $, not every $ \boldsymbol{\Psi} $ can be traced back to an allocation rule. $ \boldsymbol{\Psi} $ and $ \boldsymbol{\pi} $ that correspond to fair allocation rules satisfy at least the following constraints. First, the total allocation rate has an upper bound associated with the rate of organ arrivals relative to the rate of patient arrivals:
\begin{equation} \label{eq: constr_total}
\int_{[0,T]}d\boldsymbol{\Psi}_{s}\mathbf{1}_n \le \frac{\rho}{\tau},
\end{equation}
where $ \mathbf{1}_n $ is the vector of length $ n $ in which all elements equal to $ 1 $.
Essentially, the left side of \eqref{eq: constr_total} is the limiting proportion of patients who receive transplantation.

Second, let $\{\cdot \}_i$ denote the $i$th element of a vector, then for all $s \in [0, T]$ and $i \in \mathcal{X}$, 
\begin{prop} \label{eq: constr_bounds}
\normalfont
\begin{enumerate} 
	\item $\pi_{i}(s)\in[0,\pi_{i}(s-)]$, $d\pi_{i}(s)\in(-\infty,\{\boldsymbol{\pi}_{s-} \mathbf{Q}_s ds\}_{i}]$,
	\item $\Psi_{i}(\{s\})\in[0,\pi_{i}(s-)]$, $d\Psi_{i}(s) \in [0,\infty)$,
	\item $d\pi_{i}(s)\in[0,\{\boldsymbol{\pi}_{s-} \mathbf{Q}_s ds\}_{i}]$, $d\Psi_{i}(s)\in[0,\{\boldsymbol{\pi}_{s-} \mathbf{Q}_s ds\}_{i}]$, if $\pi_{i}(s-)=0$.
\end{enumerate}
\end{prop}

An upper bound for long-term average expected life is given by maximizing \eqref{eq: expected life gain} with respect to $ \boldsymbol{\Psi} $ and $ \boldsymbol{\pi} $ subject to \eqref{eq: constr_total} and Proposition \ref{eq: constr_bounds}.

\subsection{The Optimal Allocation Rate}

This section focuses on the characterization of the optimal $ \boldsymbol{\pi} $ and $ \boldsymbol{\Psi} $ that maximizes the objective while satisfying the constraints. The primal-dual framework is used here to convert the constrained primal problem to an unconstrained dual problem, in which a penalty parameter $c$ associated with the constraint on the limiting proportion of transplanted patients is introduced. 
Then we apply the Hamilton-Jacobi-Bellman equations to the unconstrained problem and recursively solve for the optimal transplantation rate given any value of $ c $. Dual-primal duality can be shown for this problem given the convexity of the objective and the constraints so that the primal problem is ultimately solved by finding the $ c $ corresponding to the bound in the constraint. 

Consider the unconstrained objective 
\begin{equation} \label{eq: obj_penalized}
\int_{[0,T]}d\boldsymbol{\Psi}_{s} \tilde{\boldsymbol{\upmu}}_s 
- c \int_{[0,T]}d\boldsymbol{\Psi}_{s} \mathbf{1}_n
\end{equation}
with a penalty parameter $ c \ge 0 $.
Denote the $ \boldsymbol{\Psi} $ that maximizes \eqref{eq: obj_penalized} while satisfying the boundedness constrains in Proposition \ref{eq: constr_bounds} by $\boldsymbol{\Psi}^{c}$. The following result characterizes $\boldsymbol{\Psi}^{c}$ in a recursive manner.

\begin{thm} \textbf{\nopagebreak} \label{thm: soln}
\normalfont
For all $s \in [0,T]$ and $i \in \mathcal{X}$,
\begin{equation} \label{eq: soln prodi cont}
\Psi_{i}^{c}(\{s\})
= I(\varphi_{i}^{c}(s)>c)\cdot \pi_{i}(s-)
\end{equation}
if $\pi_{i}(s-)>0$, and 
\begin{equation} \label{eq: soln when pi is zero}
d\Psi_{i}^{c}(s) 
= I(\varphi_{i}^{c}(s)>c)\cdot\sum_{j\in\mathcal{X}}\pi_{j}(s-)q_{ji}(s) ds
\end{equation}
if $\pi_{i}(s-)=0$.
Here 
\begin{equation} \label{eq: indicator phi}
\varphi_{i}^{c}(s)=\frac{\tilde{\mu}_{i}(s) - \eta_{i}^{c}(s+)}{1-\gamma_{i}^{c}(s+)},
\end{equation}
where $ \eta_{i}^{c}(s+) $ is the $ i $th element of the vector
\begin{equation} \label{eq: gamma eta}
\boldsymbol{\upeta}^{c}(s+)=\int_{(s,T]}\Prodi_{(s,t)}(\boldsymbol{I}+d\mathbf{Q})(\boldsymbol{I}-d\boldsymbol{\Lambda}^{c})\cdot d\boldsymbol{\Lambda}^{c}_{t} \tilde{\boldsymbol{\upmu}}_t,
\end{equation}
and $ \gamma_{i}^{c}(s+) $ is the $ i$th element of the vector
\begin{equation} \label{eq: gamma}
\boldsymbol{\upgamma}^{c}(s+)=\int_{(s,T]}\Prodi_{(s,t)}(\boldsymbol{I}+d\mathbf{Q})(\boldsymbol{I}-d\boldsymbol{\Lambda}^{c})\cdot d\boldsymbol{\Lambda}^{c}_{t} \mathbf{1}_n,
\end{equation}
where $ d\boldsymbol{\Lambda}^c_s $ is a diagonal matrix with the $i$th diagonal element defined as
$ d\Lambda_{i}^c(s)=d\Psi^c_{i}(s)/\pi_{i}(s-) $ if $ \pi_i(s-) \neq 0$ and
$ \Lambda^c_{i}(\{s\})=d\Psi^c_{i}(s)/\{\boldsymbol{\pi}_{s-} d\mathbf{Q}_{s}\}_{i} $ if $ \pi_i(s-) = 0 $. The symbol $ \prodi $ indicates the product integral operator as defined in \cite{Gill:1990fz}. 
\end{thm}

\begin{rem*}
Using the product integral $ \prodi $ here enables a unified expression of the transition probability of the Markov process, whether $\boldsymbol{\Psi}$ is absolutely continuous or singular, or a mixture of both. For details on the application of $ \prodi $ to the waiting list with allocations, see \cite{Zou:2015ur}.
\end{rem*}

Here is a heuristic interpretation of Theorem \ref{thm: soln}.
Imagine a scenario in which optimal allocation for a value $\rho$ of the organ arrival rate has been in place. The optimal allocation is characterized by $ d\boldsymbol{\Lambda}^c $ where $ c $ is associated with the allocation allowance $ \rho $.
Suppose we are now allowed an increase in the allocation rate to $\rho+d\rho$.
It is shown in the Appendix that $\eta^c_i(s+)$ is in fact the conditional expected life gain for a patient in state $i$ at time $s$ past arrival, if not transplanted immediately but subject to future transplantation, and $\gamma^c_i(s+)$ is the corresponding probability of future allocation.
If we choose to apply our $d\rho$ of allocation to subjects reaching state $i$ at time $s$, we will further free up $\gamma^c_i(s+) d\rho$ more allocation that could be applied at state $i$ and time $s$, further freeing up $(\gamma^c_i(s+))^2 d\rho$, $\dots$ 
Ultimately, there would be a $d\rho/(1-\gamma^c_i(s+))$ increase in allocation at state $i$ and time $s$ and an expected life again of $ \tilde{\mu}_i(s) - \eta^c_i(s+) $ per unit of increase in allocation.
This suggests that optimal allocation occurs at those $i$ and $s$ such that
\begin{equation*}
\varphi_{i}^{c}(s)=\frac{\tilde{\mu}_{i}(s) - \eta^c_{i}(s+)}{1-\gamma^c_{i}(s+)}
\end{equation*}
is large.

\section{Allocation Strategy} \label{sect: strategy}
Theorem \ref{thm: soln} suggests an upper bound of the long-term average of the expected life gain can be achieved by allocating organs to patients in state $ i $ at time $ s $ if and only if $ \varphi_{i}^{c}(s) > c $, where $ c $ is selected such that $\int_{[0,T]}d\boldsymbol{\Psi}^{c}_{s}\mathbf{1}_n = \rho / \tau$.

In reality, however, this upper bound cannot be reached, as when an organ is available for transplantation, there might not be any patient in the optimal state and waiting time in the current waiting list. Since the organ can only be preserved for a limited time before losing its functionality, the transplantation cannot be delayed for patients in the optimal state and waiting time to appear. One of the patients available in the waiting list, though in suboptimal states and waiting times, has to be selected for transplantation. Therefore, an allocation strategy is needed to prioritize patients in all possible states and waiting times.

Here we propose an allocation strategy motivated by the form of the optimal transplantation rate. The proposed strategy are developed based on three monotonicity results. 
First, the total transplantation rate is a monotone function of the penalty parameter $ c $.
\begin{prop} 
\normalfont
Suppose $ c_1 > c_2 \ge 0 $, then 
\begin{equation}
\int_{[0,T]} d\boldsymbol{\Psi}^{c_1}_{s} \mathbf{1}_n \le \int_{[0,T]} d\boldsymbol{\Psi}^{c_2}_{s} \mathbf{1}_n.
\end{equation}
\end{prop}

Second, the long-term average expected life gain from transplantation is a monotone function of $c$.
\begin{prop}
\normalfont
Suppose $ c_1 > c_2 \ge 0 $, then
\begin{equation}
\int_{[0,T]}d\boldsymbol{\Psi}^{c_1}_{s} \tilde{\boldsymbol{\upmu}}_s 
\le
\int_{[0,T]}d\boldsymbol{\Psi}^{c_2}_{s} \tilde{\boldsymbol{\upmu}}_s. 
\end{equation}
\end{prop} 

Third, the optimal set of state and waiting time pairs for transplantation is monotone with respect to $ c $.
Formally, for each value of $c \ge 0$, define the corresponding allocable set 
\begin{equation} \label{eq: feasible}
\mathcal{A}_{c}= \{(i,s) \in \mathcal{X} \times [0,T]: \varphi^c_i(s) > c \},
\end{equation}
which consists of state and waiting time pairs with non-zero transplantation rate in the optimal solution given by Theorem \ref{thm: soln}. The following result states that $ \mathcal{A}_c $ expands monotonically as $ c $ decreases.

\begin{thm} \label{thm: monotone}
\normalfont
Suppose $ c_1 > c_2 \ge 0 $, then
\begin{equation} \label{eq:cont monotonicity of allocation rule}
\mathcal{A}_{c_1} \subseteq \mathcal{A}_{c_2}.
\end{equation}
\end{thm}

Based on the monotonicity results, a full order priority ranking of all state and waiting time pairs can be obtained with the following algorithm outlined in the pseudo code Algorithm \ref{alg: rank} below:
start with a large enough value of $ c $ such that $\mathcal{A}_{c} = \emptyset$,
then the value of $ c $ is gradually decreased, representing increasing organ availability, and the order in which each state and waiting time pair enter the allocable set $ \mathcal{A}_c $ as $c$ decreases is recorded. A binary search is used to find each value of $c$ when a new pair of $(i,s)$ enters $ \mathcal{A}_c $.
Pairs that enter the allocable set earlier are given higher priorities for transplantation, as they are selected for allocation to prolong the long-term average expected total life when the availability of organs is more stringent and remain allocable when more organs are available. 
The sensitivity parameter $\alpha$ in Algorithm \ref{alg: rank} is a small positive number needed in the algorithm to avoid infinite loops. Smaller $\alpha$ leads to higher precision in ranking all waiting time and state combinations. $N_{\text{steps}}$ is the maximal number of steps allowed in calculating the ranks of $(i,s)$ combinations.

\begin{algorithm}
\caption{Priority Ranking for Transplantation}\label{alg: rank}
\begin{algorithmic}[1]
\Procedure{Rank}{$\{(i,s): i \in \mathcal{X}, s \in [0, T]\}$}
\State Select $ c > 0$ 			\Comment{Start with an arbitrary value of $c$}
\If {$\mathcal{A}_c \neq \emptyset$} 			\Comment{Increase $c$ to ensure $\mathcal{A}_c = \emptyset$}
\While {$\mathcal{A}_c \neq \emptyset$}
\State {$c \leftarrow c +1$}
\EndWhile
\EndIf
\State $c_0 \leftarrow c$; $L \leftarrow 0$		
\Comment{Define upper and lower bounds for $c$}
\For{step = $1$ to $N_{\text{steps}}$ } 		
\State $c_u \leftarrow c_{\text{step}-1}$; $c_l \leftarrow L$; $c_{\text{prev}} \leftarrow L$; $c \leftarrow (c_u + c_l)/2$ 
\Comment{Initialize binary search}
\While{$|c_{\text{prev}} - c| > \alpha$} 
\Comment{Find largest $c < c_{\text{step}-1}$ }
\If {$\mathcal{A}_c \setminus \mathcal{A}_{c_{\text{prev}}} \neq \emptyset$}			
\Comment{s.t. $\mathcal{A}_c \setminus \mathcal{A}_{c_{\text{step}-1}}\neq \emptyset$}
\State $c_l \leftarrow c$; $c_{\text{prev}} \leftarrow c$; $c \leftarrow (c_u + c_l)/2 $
\Else 
\State $c_u \leftarrow c$; $c \leftarrow (c_u + c_l)/2 $
\EndIf
\EndWhile
\If {$\mathcal{A}_c \setminus \mathcal{A}_{c_{\text{prev}}} = \emptyset$}
\State $c \leftarrow c_{\text{prev}} $
\EndIf
\State $c_{\text{step}} \leftarrow c$; $\text{rank}(i,s) \leftarrow $ step for all $ (i,s) \in \mathcal{A}_{c_{\text{step}}} \setminus \mathcal{A}_{c_{\text{step}-1}} $ 		
\EndFor
\State \textbf{return} rank

\EndProcedure
\end{algorithmic}
\end{algorithm}

Once the rank of state and waiting time pairs is determined, whenever an organ is retrieved, patients available in the current waiting list can be instantly ranked based on their state and waiting time. 
The patient with the highest rank will be selected for transplantation. 
The resulting long-term average expected total life using this strategy is expected to be close to the upper bound, as patients are ranked according to their potential contributions to the limiting average total life considering the limited availability of organs. This allocation strategy satisfies the fairness requirement, as the allocation decisions are solely determined by states and waiting times of patients in the current list.

\section{Matching of Multiple Organ Types} \label{sect: multi}

Now we generalize the allocation problem with single organ type to multiple organ types.
Suppose there are $N_o> 1$ types of organs with corresponding arrival rates $\{\rho^{(w)}: w = 1, \dots, N_o  \}$. 
The interesting case is when $\sum_{w=1}^{N_o} \rho^{(w)} < \tau$ which reflects the reality that there are not enough organs to meet the need of the patients.

For any $ w  \in \{1, \dots, N_o \}$, let $\tilde{\upmu}_{i}^{(w)}(s)$ denote the expected gain in residual life for a patient in state $i$ and waiting time $s$ with immediate transplantation with a type $ w $ organ comparing to never transplanted. 
Note $\tilde{\upmu}_{i}^{(w)}(s)$ is a function of both the state and waiting time, providing the flexibility to model the benefit of transplantation with each organ type as depending on the compatibility of the organ with the patient and possibly varying with waiting time.
In addition to factors that are usually taken into account in matching donors and recipients such as blood type, patients' health states can include factors that are found to be significantly related to the benefits of different organ types. 
For example, studies (\cite{Oto:2004dr}, \cite{Bonser:2012ix}, \cite{Shigemura:2013ej} and \cite{Kopp:2015uq}) have found that while the post-transplant survival is worse with organs from donors with a smoking history, certain recipient characteristics could make a difference in the benefit of transplantation with a smoking-donor organ.

\subsection{The Optimal Transplantation Rate for Multiple Organ Types}
It can be shown, as in the single organ type case, that with any allocation rule that satisfies fairness requirements for each organ type there exists a limiting transplantation rate $\boldsymbol{\Psi}^{(w)}_s$ for each $ w  = 1, \dots, N_o$, with which the limiting averaged life gain can be represented by
\begin{equation*}
\sum_{w=1}^{N_o}\int_0^T d\boldsymbol{\Psi}_{s}^{(w)} \tilde{\boldsymbol{\upmu}}_s^{(w)}.
\end{equation*}
Moreover, for each $ w $, the constraint
\begin{equation} \label{eq: multi_organ_constraints}
\int_{[0,T]}d\boldsymbol{\Psi}_{s}^{(w)} \mathbf{1}_n \le \frac{\rho^{(w)}}{\tau},
\end{equation}
must hold as the limiting proportion of patients transplanted with organ type $ w $ is bounded by the ratio of type $ w $ organ arrival rate to the patient arrival rate.
Each $\boldsymbol{\Psi}_{s}^{(w)}$ must also satisfy the first condition in Proposition \ref{eq: constr_bounds} and the following constraints:
\begin{equation} \label{eq: mult organ constraint 1}
\sum_{w=1}^{N_o} \Psi_{i}^{(w)}(\{s\})  \in [0,\pi_{i}(s-)] \text{ and } \sum_{w=1}^{N_o} d\Psi_i^{(w)}(s) \in [0, \infty), \text{ if } \pi_i(s-) \neq 0,
\end{equation}
and 
\begin{equation} \label{eq: mult organ constraint 2}
\sum_{w=1}^{N_o}  d\Psi_{i}^{(w)}(s)\in[0,\{\boldsymbol{\pi}_{s-} d\mathbf{Q}_s\}_{i}], 
\text{ if } \pi_i(s-) = 0, 
\end{equation}
as a patient can only be transplanted with one organ type at a time.

As in the single organ type case, the optimal allocation rates can be solved by first considering the penalized objective
\begin{equation} \label{eq: multi organ obj}
\sum_{w=1}^{N_o} 
\left(
\int_{[0,T]}d\boldsymbol{\Psi}^{(w)}_{s} \tilde{\boldsymbol{\upmu}}_s 
- c^{(w)} \int_{[0,T]}d\boldsymbol{\Psi}^{(w)}_{s} \mathbf{1}_n
\right),
\end{equation}
where each $c^{(w)} \geq 0$ is a penalty parameter associated with the constraint \eqref{eq: multi_organ_constraints}, indicating the availability of type $ w $ organs.
Let $\mathbf{c} = (c^{(1)},\dots, c^{(N_o)})$ denote the vector of penalty parameters for all organ types.
The following result characterizes the optimal allocation rates for \eqref{eq: multi organ obj} given any fixed value of $\mathbf{c}$. 

\begin{thm} \textbf{\nopagebreak} \label{thm: soln_multi}
	\normalfont
	For all $s \in [0,T]$ and $i \in \mathcal{X}$, the transplantation rates $\{\Psi_{i}^{(w),\, c}: w = 1, \dots, N_o \}$ that maximize \eqref{eq: multi organ obj} are given by
	\begin{multline} \label{eq: soln prodi cont_m}
	\Psi_{i}^{(w),\,c}(\{s\})
	= I \Big(\tilde{\upmu}^{(w)}_i(s) - c^{(w)} > 
	\sum_{w=1}^{N_o}  \big[ \eta_{i}^{(w),\,c}(s+) - c^{(w)} \gamma_{i}^{(w),\,c}(s+) \big] ,\\
	\tilde{\upmu}^{(w)}_i(s) - c^{(w)}  > \tilde{\upmu}^{(v)}_i(s) - c^{(v)}, \, \forall v \neq w
	\Big)
	\cdot \pi_{i}(s-)
	\end{multline}
	if $\pi_{i}(s-)>0$, and 
	\begin{multline} \label{eq: soln when pi is zero_m}
	d\Psi_{i}^{(w),\,c}(s) 
	= I \Big(\tilde{\upmu}^{(w)}_i(s) - c^{(w)} > 
	\sum_{w=1}^{N_o}  \big[ \eta_{i}^{(w),\,c}(s+) - c^{(w)} \gamma_{i}^{(w),\,c}(s+) \big] ,\\
	\tilde{\upmu}^{(w)}_i(s) - c^{(w)}  > \tilde{\upmu}^{(v)}_i(s) - c^{(v)}, \, \forall v \neq w
	\Big)
	\cdot\sum_{j\in\mathcal{X}}\pi_{j}(s-)q_{ji}(s) ds
	\end{multline}
	if $\pi_{i}(s-)=0$,
	where $ \eta_{i}^{(w),\,c}(s+) $ is the $ i $th element of the vector
	\begin{equation} \label{eq: gamma eta_m}
	\boldsymbol{\upeta}^{(w),\,c}(s+)=\int_{(s,T]}\Prodi_{(s,t)}(\boldsymbol{I}+d\mathbf{Q})
	(\boldsymbol{I}- \sum_{l=1}^{N_o} d\boldsymbol{\Lambda}^{(l),\,c})\cdot d\boldsymbol{\Lambda}^{(w),\,c}_{t} \, \tilde{\boldsymbol{\upmu}}^{(w)}_t,
	\end{equation}
	and $ \gamma_{i}^{(w),\,c}(s+) $ is the $ i$th element of the vector
	\begin{equation} \label{eq: gamma_m}
	\boldsymbol{\upgamma}^{(w),\,c}(s+)=\int_{(s,T]}\Prodi_{(s,t)}(\boldsymbol{I}+d\mathbf{Q})
	(\boldsymbol{I}- \sum_{l=1}^{N_o} d\boldsymbol{\Lambda}^{(l),\,c})\cdot d\boldsymbol{\Lambda}^{(w),\,c}_{t} \, \mathbf{1}_n,
	\end{equation}
	where each $ d\boldsymbol{\Lambda}^{(w),\,c}_s $ is a diagonal matrix with the $i$th diagonal element being
	$ d\Lambda^{(w),\,c}_{i}(s)=d\Psi^{(w),\,c}_{i}(s)/\pi_{i}(s-) $ if $ \pi_i(s-) \neq 0$ 
	and
	$ \Lambda^{(w),\,c}_{i}(\{s\})=d\Psi^{(w),\,c}_{i}(s)/\{\boldsymbol{\pi}_{s-} d\mathbf{Q}_{s}\}_{i} $ if $ \pi_i(s-) = 0 $. 
\end{thm}

Theorem \ref{thm: soln_multi} indicates that the optimal allocation rates for multiple organ types in the heavy traffic setting can be achieved by allocating all of patients' occupancy in state $i$ at waiting time $s$ to the transplanted state with type $ w $ organs if both of the following conditions are satisfied: 
first, the difference between the post-transplant residual life gain and the organ-type-specific penalty, if immediately transplanted, is the largest when transplanted with a type $ w $ organ comparing to other organ types, which makes type $ w $ the ``designated'' organ type for $(i,s)$;
and second, the difference between the post-transplant residual life gain and the organ-type-specific penalty parameter, if transplanted with a type $ w $ organ immediately, is larger than the expected difference between the residual life gain and the penalty parameter if the patient is not transplanted immediately but open to future transplantation opportunities with any type of organ.

\subsection{Allocation Strategy for Multiple Organ Types} \label{sec: multi organ strategy}

In practice, as in the single organ type case, the optimal transplantation rates usually cannot be reached, as when a type $ w $ organ becomes available, there may not be any patient in the optimal state and waiting time for this specific organ type in the waiting list. Consequently, an allocation strategy is needed to rank patients' priorities of transplantation with each organ type.  

Theorem \ref{thm: soln_multi} motivates a natural allocation strategy.
Given the form of the optimal transplantation rate, for each $ w \in \{1, \dots, N_o\}$, the allocable set of state and waiting time combinations for type $ w $ organs is denoted by
\begin{align*}
\mathcal{A}^{(w)}_{\mathbf{c}} 
= \big\{(i,s) \in \mathcal{X}\times [0,T]:  
&\,\tilde{\upmu}^{(w)}_i(s) - c^{(w)} > \sum_{w=1}^{N_o}  
\big[ \eta_{i}^{(w),\,c}(s+) - c^{(w)} \gamma_{i}^{(w),\,c}(s+) \big] ,\\
&\,\tilde{\upmu}^{(w)}_i(s) - c^{(w)}  > \tilde{\upmu}^{(v)}_i(s) - c^{(v)}, \, \forall v \neq w
\big\},
\end{align*} 
in which $\mathbf{c}$ indicates the availability of each type of organs.
If each $\mathcal{A}^{(w)}_{\mathbf{c}} $ expands monotonically when $\mathbf{c}$ decreases to reflect more organs becoming available, then a priority ranking for transplantation with each organ type can be obtained by comparing the order of each $(i,s)$ entering the $\{\mathcal{A}^{(w)}_{\mathbf{c}}\}$ as $\mathbf{c}$ decreases. Pairs that enter $\mathcal{A}^{(w)}_{\mathbf{c}}$ earlier are given higher priorities for transplantation with type $ w $ organ.

The monotonicity of $\{\mathcal{A}^{(w)}_{\mathbf{c}} \}$, however, generally does not hold when $\mathbf{c}$ decreases freely.  
In fact, the monotonicity in the multiple organ type case only holds in a weaker sense, that is, when the decreasing of $\mathbf{c}$ follows special ``paths'' with constraints on changes in each $c^{(w)}$.
The following results describe two types of weak monotonicity properties.
First, when all of the $\{c^{(w)}\}$ decrease by the same amount, the allocable set for each organ type expands monotonically. 
Second, when fixing all the $c^{(v)}$ where $v \neq w$ and decreasing only $c^{(w)}$, the allocable set for type $ w $ organs expands monotonically. 

\begin{prop} \label{thm: weak monotone 1}
	\normalfont	
	If $c^{(w)}_1 - c^{(w)}_2$ is a positive constant independent of $ w $, 
	then $\mathcal{A}^{(w)}_{\mathbf{c}_1} \subseteq \mathcal{A}^{(w)}_{\mathbf{c}_2}$ for all $ w  \in \{1, \dots, N_o\}$.				
\end{prop}

\begin{prop} \label{thm: weak monotone 2}
	\normalfont
	If $ c^{(w)}_1 > c^{(w)}_2 $ and $c^{(v)}_1 = c^{(v)}_2 $ for all $v \neq w$, 
	then $\mathcal{A}^{(w)}_{\mathbf{c}_1} \subseteq \mathcal{A}^{(w)}_{\mathbf{c}_2}$.
\end{prop}

Following the above paths of decreasing $\mathbf{c}$, the priority rankings for all organ types can be obtained using a three-stage algorithm:
\begin{enumerate}
	\item The value of $\mathbf{c}$ such that the active constraints \eqref{eq: multi_organ_constraints} are satisfied, at least approximately so, are calculated via a multiple variable gradient descent algorithm.  This step guarantees that $\mathbf{c}$ reflects the practical relative availability of each organ type.\\
	
	\item Given the calculated value of $\mathbf{c}$, increase $\mathbf{c}$ by a constant so that the new $\mathcal{A}^{(w)}_{\mathbf{c}}$ is empty for each $ w $. Then decrease $\mathbf{c}$ by the same amount as in Proposition \ref{thm: weak monotone 1} and rank the priorities for type $ w $ organ transplantation according to the order of each $(i,s)$ pair entering $\mathcal{A}^{(w)}_{\mathbf{c}}$. 
	
	When decreasing each $c^{(w)}$ by the same amount, the designated organ type $ w $ for each $(i,s)$ that maximizes $\tilde{\upmu}^{(w)}_i(s) - c^{(w)} $ remains the same. Consequently, when $\mathbf{c}$ is small enough for each $(i,s)$ to be allocable, each $(i,s)$ has a designated organ type and the space of $\mathcal{X} \times [0,T]$ is partitioned into disjoint subsets indexed by the designated organ types. \\
	
	\item For each $ w $, decrease $c^{(w)}$ further while keeping all other $c^{(v)}, v \neq w$ constant as in Proposition \ref{thm: weak monotone 2}, until all $(i,s) \in \mathcal{X} \times [0,T]$ are ranked for type $ w $ organs. This step guarantees that for each organ type there is a complete ranking of combinations in $\mathcal{X} \times [0,T]$.
	This step is necessary since in practice when an organ of type $ w $ arrives, there might not be any patient in the subset of $\mathcal{X} \times [0,T]$ with type $ w $ as the designated organ type. An allocation decision still needs to be made, since there cannot be a delay in transplantation due to the time limit for the organ to keep its functionality, and patients who are currently in the waiting list might still benefit from the transplantation. 
	Complete rankings of $\mathcal{X} \times [0,T]$ for each organ type ensures an allocation decision can be made to select the patient who would benefit the most when an organ becomes available.
\end{enumerate}

\section{Comparison of Allocation Strategies} \label{sect: comparison}

This section focuses on the application of the proposed allocation strategy in the single-organ-type scenario in the context of realistic models for health state transitions of patients in the lung transplantation waiting list. Waiting lists are simulated based on parameters estimated from the United Network for Organ Sharing (UNOS) lung transplantation data, and the proposed strategy is compared to the Lung Allocation Score (LAS) system by its performance in extending life when applied to the simulated waiting lists. 

\subsection{The United Network for Organ Sharing (UNOS) Data}

The data used are the waiting list, transplant and follow-up UNOS Standard Transplant Analysis and Research files, which contain information of heart, lung, and simultaneous heart-lung registrations and transplants that were listed or performed in the United States and reported to the Organ Procurement and Transplantation Network (OPTN) from October 1, 1987 to December 31, 2012. In the data, the first transplantation with the LAS occurred on May 5, 2005. Before June 11, 2013, only patients at least 12 years of age received priority for deceased donor lung offers based on the LAS, and thus we only include patients at least 12 years of age in the data analysis.
There were 16,049 such patients registered in the above time period, with 129,881 records of medical measurements updated sporadically at different times for different patients during their tenure on the waiting list. Among the 16,049 candidates, 64.6\% received transplantation, 18.4\% died while waiting in the list, and the remaining were still waiting at the end of the study period or censored by loss of follow-up. Among patients who received transplantation, 37.8\% died and 62.2\% were still alive at the end of the study period or were censored due to loss of follow-up.

\subsection{Parameter Estimation from the UNOS Data}

We fit two separate proportional hazards regression models with time varying covariates to estimate the time-on-waiting-list specific hazards for in-waiting-list and post-transplant deaths. This serves two purposes: 1. to calculate the LAS given a patient's current covariates values and 2. to characterize patients' health states. 

Covariates in the proportional hazard regressions are those used in the calculations in the revised LAS system by UNOS (for details of covariates used in the LAS calculation, see \cite{UNOS:2015wn}), whenever they are available in the data, so that no advantage is gained from an improved variable selection when comparing the proposed strategy to the LAS. 
Note in estimating the hazard for in-waiting-list death, transplantation leads to censoring in the data. The censoring by transplantation can be treated as censoring at random, nevertheless, as the selection of organ recipients was based on patients' LAS calculated with the same covariates used in estimating the hazard for in-waiting-list death and thus can be treated as conditionally independent of future survivals given the observed covariates. 
See Tables \ref{t1} and \ref{t2} in the Appendix for lists of covariates and coefficient estimates in the two proportional hazards regressions.
More details including estimated baseline survival curves can be found in the Supplementary Material.

\subsection{Defining Patients' Health States} \label{subsec: health states}

Patients' health states are characterized in terms of the linear combinations of the covariates in the proportional hazards models, with the estimated regression coefficients being the weights associated with the covariates. 
Each patients' health state at waiting time $ s $ is defined as the pair $(S_{wl}(s),S_{\mu}(s))$, where $ S_{wl}(s) $ and $ S_{\mu}(s) $ are, respectively, values of the linear combinations of covariates at $ s $ in the proportional hazards models for in-waiting-list and post-transplant survivals.

% transition
Characterizations of transitions in health states are based on observations from exploratory analysis of the data. Covariates relevant to waiting list and post-transplant survivals are categorized into deterministic and stochastic variables. 
Deterministic variables are those remain mostly constant or are deterministic functions of waiting time. Covariates such as age, diagnosis group and detailed diagnosis fall into this category. 
Stochastic variables are those change randomly while patients are waiting for transplants. Examples of stochastic variables include BMI, functional assistance status, ventilation status, creatinine, oxygen and six minute walk distance.

In the LAS calculation (\cite{UNOS:2015wn}), the set of the stochastic covariates for estimating post-transplant survival is a subset of the stochastic covariates used in estimating the in-waiting-list survival. Therefore, the set of stochastic covariates can be partitioned further into two sub-categories: covariates used in both proportional hazards models for in-waiting-list survival and post-transplant survival, and covariates used only for in-waiting-list survival. As a result, the linear combinations $ S_{wl}(s) $ and $ S_{\mu}(s) $ can be written as
\begin{align} \label{eq: defn S}
S_{wl}(s) & = \boldsymbol{\upbeta}_1 \cdot \mathbf{X}_{1}(s) + \boldsymbol{\upbeta}_2 \cdot \mathbf{X}_{2}(s) +  \boldsymbol{\upbeta}_3 \cdot \mathbf{X}_{3}(s), \\
S_{\mu}(s) & = \tilde{\boldsymbol{\upbeta}}_{1} \cdot \mathbf{X}_{1}(s) + \boldsymbol{\upbeta}_4 \cdot \mathbf{X}_{4}(s), \nonumber 
\end{align}
where $ \mathbf{X}_{1} $ represents stochastic covariates included in both models, $ \mathbf{X}_{2} $ represents stochastic covariates for $ S_{wl} $ only, $\mathbf{X}_{3} $ represents deterministic covariates for $ S_{wl} $, and $ \mathbf{X}_{4} $ represents deterministic covariates for $ S_{\mu} $. 
See Tables \ref{t1} and \ref{t2} in the Appendix for detailed categorizations of the covariates into each of the five $ \mathbf{X}$ categories.

\begin{figure}[H]
	\centering
	\includegraphics[width=0.5\linewidth]{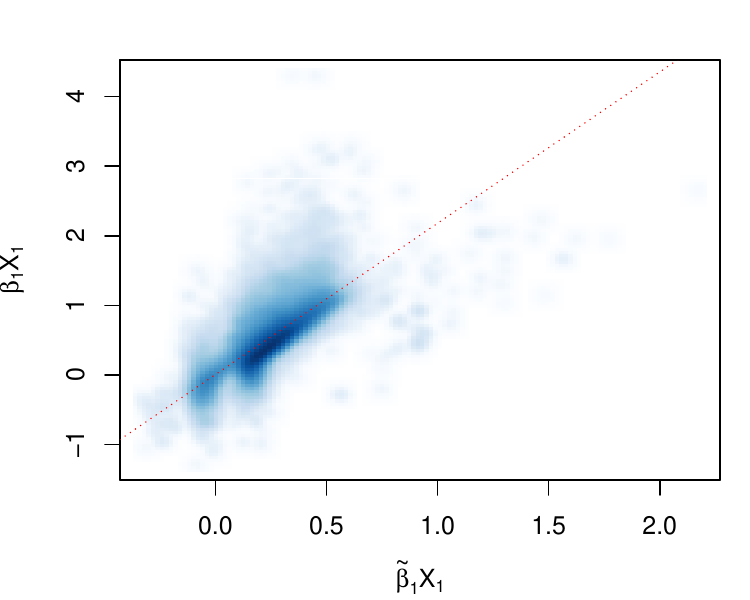}
	\caption{Linear Dependence Between $ \boldsymbol{\upbeta}_1 \mathbf{X}(s) $ and $ \tilde{\boldsymbol{\upbeta}}_1 \mathbf{X}(s)$} \label{fig:scatterbetax}
\end{figure}

$ \boldsymbol{\upbeta}_1 \mathbf{X}_{1}(s) $ and $ \boldsymbol{\upbeta}_2 \mathbf{X}_{2}(s) $ are characterized as following conditionally independent compound jumping processes whose jumping intensities and magnitudes depend on the current waiting time and health state $(S_{wl},S_{\mu})$. The jumping hazards are estimated with proportional hazards regressions for recurrent events, while the jumping magnitudes (in logarithmic scale) are estimated with linear regressions. 
It was observed from the data that estimates of corresponding coefficients in $ \boldsymbol{\upbeta}_1 $ and $ \tilde{\boldsymbol{\upbeta}}_1 $ are approximately proportional (Figure \ref{fig:scatterbetax}). Therefore the transition in $ \tilde{\boldsymbol{\upbeta}}_1 \mathbf{X}_{1}(s) $ can be approximated by the transition in $ \boldsymbol{\upbeta}_1 \mathbf{X}_{1}(s) $ scaled by the estimate of their ratio, and vice versa.

The transition probabilities in $(S_{wl}(s),S_{\mu}(s))$ can be derived from the transition probabilities of $ \boldsymbol{\upbeta}_1\mathbf{X}_{1}(s) $ and $ \boldsymbol{\upbeta}_2 \mathbf{X}_{2}(s) $ and the deterministic trajectories of $ \boldsymbol{\upbeta}_3\mathbf{X}_{3}(s) $ and $ \boldsymbol{\upbeta}_4 \mathbf{X}_{4}(s) $. Specifically, for any $U, V \subset \mathbb{R}$,
\begin{align} \label{eq: convolution}
& P(S_{wl}(s+\Delta s) \in U, S_{\mu}(s+ \Delta s) \in V \mid S_{wl}(s) = x, S_{\mu}(s) = y)  \\
& = \int_{V_{\Delta \boldsymbol{\upbeta}_4 \mathbf{X}_{4}(s)}} P(\Delta \boldsymbol{\upbeta}_{2} \mathbf{X}_{2}(s) \in U_{l(\tilde{y}) + \Delta \boldsymbol{\upbeta}_3 \mathbf{X}_{3}(s)} \mid S_{wl}(s) = x, S_{\mu}(s) = y) \nonumber\\
& \qquad \qquad \qquad \cdot dP(\Delta  \tilde{\boldsymbol{\upbeta}}_{1} \mathbf{X}_{1}(s) = \tilde{y} \mid S_{wl}(s) = x, S_{\mu}(s) = y), \nonumber
\end{align}
where $ V_x $ denotes the interval $ V $ shifted by $ -x $ and $ l(\Delta \tilde{\boldsymbol{\upbeta}}_{1}\mathbf{X}_{1}(s)) = \Delta \boldsymbol{\upbeta}_{1} \mathbf{X}_{1}(s)$ is the linear function approximating the linear relationship between the changes in $ \boldsymbol{\upbeta}_1 \mathbf{X}(s) $ and $ \tilde{\boldsymbol{\upbeta}}_1 \mathbf{X}(s)$, and each conditional probability in the right side of \eqref{eq: convolution} is calculated using the estimated jumping intensities and magnitudes described earlier.
The distribution of initial states $(S_{wl}(0), S_{\mu}(0))$ is estimated with the empirical distribution of patients' states upon arrival to the waiting list.

\subsection{Discretization of Patients' Health States} \label{subsec: discretization}

For the purpose of efficiency in computation, we discretize waiting time to periods of $30$ days and $(S_{wl}, S_{\mu})$ to a $4\times 4$ finite state space. 
Each of $S_{wl}$ and $S_{\mu}$, which were originally in continuous scales, are partitioned into $4$ intervals, where the first and last cut-off points of the intervals are the 1/4 and 3/4 empirical quantiles of the original variables defined in \eqref{eq: defn S}, and the rest of the cut-off points are such that the intervals have equal lengths. As a result there are $4$ states: $\{1, 2, 3, 4\}$ for each of $S_{wl} $ and $ S_{\mu}$, and the overall discretized patient state space is defined as the cross product of the discretized state spaces of $S_{wl} $ and $ S_{\mu}$.

The corresponding discrete transition probabilities among states are calculated from the estimated continuous transition probabilities. Specifically, at each waiting time $s$, the transition probability from state $i = (i_1, i_2)$ to $j = (j_1, j_2)$ as waiting time increases by $\Delta s$ is approximated with \eqref{eq: convolution}, in which $x$ and $y$ are the middle points (medians in case of the first and last intervals) of the $i_1$th interval of $S_{wl}$ and $i_2$th interval of $S_{\mu}$, and $U$ and $V$ are the $j_1$th interval of $S_{wl}$ and $j_2$th interval of $S_{\mu}$, respectively. We first estimate the single-day transition probabilities with $\Delta s = 1$, and then the cumulative transition probabilities for the discretized waiting time intervals with $\Delta s = 30$ are calculated as the product of the single-day transition probabilities using the Markov property.

We also experimented with different fineness levels in the discretization of waiting time and health states as well as different models for transitions in health states and simulation results were robust to different settings.
See the Supplementary Material for another example from \cite{Zou:2015ur} in which waiting time was discretized into $90$-day periods and $(S_{wl}, S_{\mu})$ was discretized into a $3 \times 3$ state space and the transition probabilities of the discretized $(S_{wl}, S_{\mu})$ was estimated with a multinomial regression model.

\subsection{Calculation of Priority Rankings of Different Allocation Strategies}

Given the defined health states, allocation strategies can be formulated as priority rankings of all combinations of waiting time and health state.
Here we compare our proposed allocation strategy with the LAS and the refined LAS-type method. 
First, the LAS currently used by the UNOS is $\textrm{LAS}=100\cdot(\textrm{PTAUC}-2\cdot\textrm{WLAUC}+730)/1095$,
where WLAUC is the estimated in-waiting-list life expectancy during an additional year and PTAUC is the estimated post-transplant life expectancy during the first year, given the patient's current waiting time and state $(S_{wl},S_{\mu})$. Patients with higher LAS have more priorities of transplantation. Specifically, the two life expectancy measures are calculate as follows:
\begin{equation*}
\text{WLAUC} = \sum_{t=0}^{364} \bar{F}_{wl}(t), \qquad \text{PTAUC} = \sum_{t=0}^{364} \bar{F}_{tx}(t),
\end{equation*}
where $\bar{F}_{wl}(t)$ is the in-waiting-list survival function at time $t$ (treating current waiting time as time $0$), and $\bar{F}_{tx}(t)$ is the post-transplant survival function at time $t$ (treating time at transplant as time $0$).

It is observed from the data that patients' in-waiting-list and post-transplant life are usually much longer than one year. 
Studies showed the emphasis on one-year survival by the LAS might have led to worse long-term survival. See for example, \cite{Maxwell:2014kf}, for detailed statistics.
Moreover, it is implicitly assumed in the LAS calculation that the in-waiting-list survival functional $\bar{F}_{wl}(t)$ is invariant to the patient's current waiting time, which may not be case. 
Here we also calculate a refined LAS without the one-year constraint on the life expectancy measures and without assuming the invariance to current waiting time. 
Due to heavy censoring of organ recipients with post-transplant residual life longer than five years, in the refined LAS we calculated the median of the covariate-specific post-transplant survival, as it is less sensitive to missing values comparing to the mean, as the post-transplant life measure.

The priority ranking of all waiting time and health state combinations in our proposed allocation strategy is calculated with the following procedure.
Given parameters estimated from the UNOS data, for a fixed value of the penalty parameter $c$, whether $(s,i)$, in which $s$ is the waiting time and $i$ is the value of the health state $(S_{wl}, S_{\mu})$, is allocable is decided by whether the optimal transplantation rate for this combination is non-zero. The optimal transplantation rate can be calculated with formula \eqref{eq: soln when pi is zero}, which in turn is decided by $\varphi_i^c(s)$ in \eqref{eq: indicator phi}.
The calculation of $\varphi_i^c(s)$ requires the estimation of $\tilde{\mu}_i(s)$, $\eta_i^c(s+)$, and $\gamma_i^c(s+)$. $\tilde{\mu}_i(s)$ is calculated as the difference between the estimated post-transplant residual life and the estimated in-waiting-list life if never transplanted, where both residual life are estimated with proportional hazards models given the current waiting time and health state. 
The estimated $\eta_i^c(s+)$ and $\gamma_i^c(s+)$ are calculated using formulae \eqref{eq: gamma eta} and \eqref{eq: gamma}, in which the counterfactual transition matrix $\mathbf{Q}$ is estimated from the data with the method described in Sections \ref{subsec: health states} and \ref{subsec: discretization} and $d\mathbf{\Lambda}^c$ is calculated backwards in an iterative manner from the upper bound $T$ of waiting time.
Finally, based on the monotonicity of the allocable sets $ \mathcal{A}_c $ in Theorem \ref{thm: monotone}, a full order ranking of transplant priorities for combinations of waiting times and states is obtained by comparing the order of their entries to $ \mathcal{A}_c $ when $c$ decreases as described in Algorithm \ref{alg: rank}. 

\begin{figure}[H]
\includegraphics[width=0.98\linewidth]{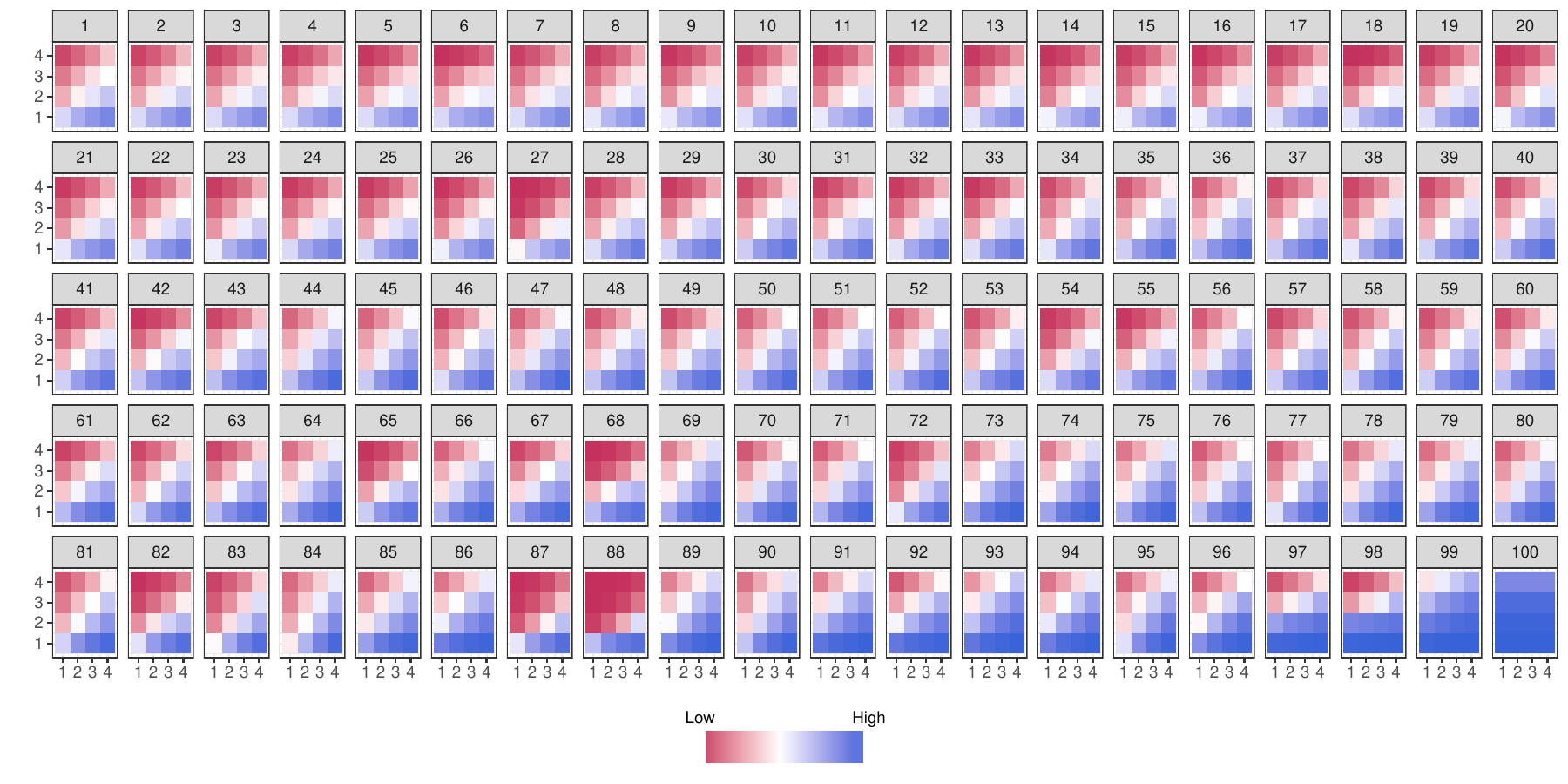} 		
\captionof{figure}{Transplantation priorities of waiting time and health state combinations with the proposed strategy  \label{Optimal}} 
\end{figure}

\begin{figure}[H]
\includegraphics[width=0.98 \linewidth]{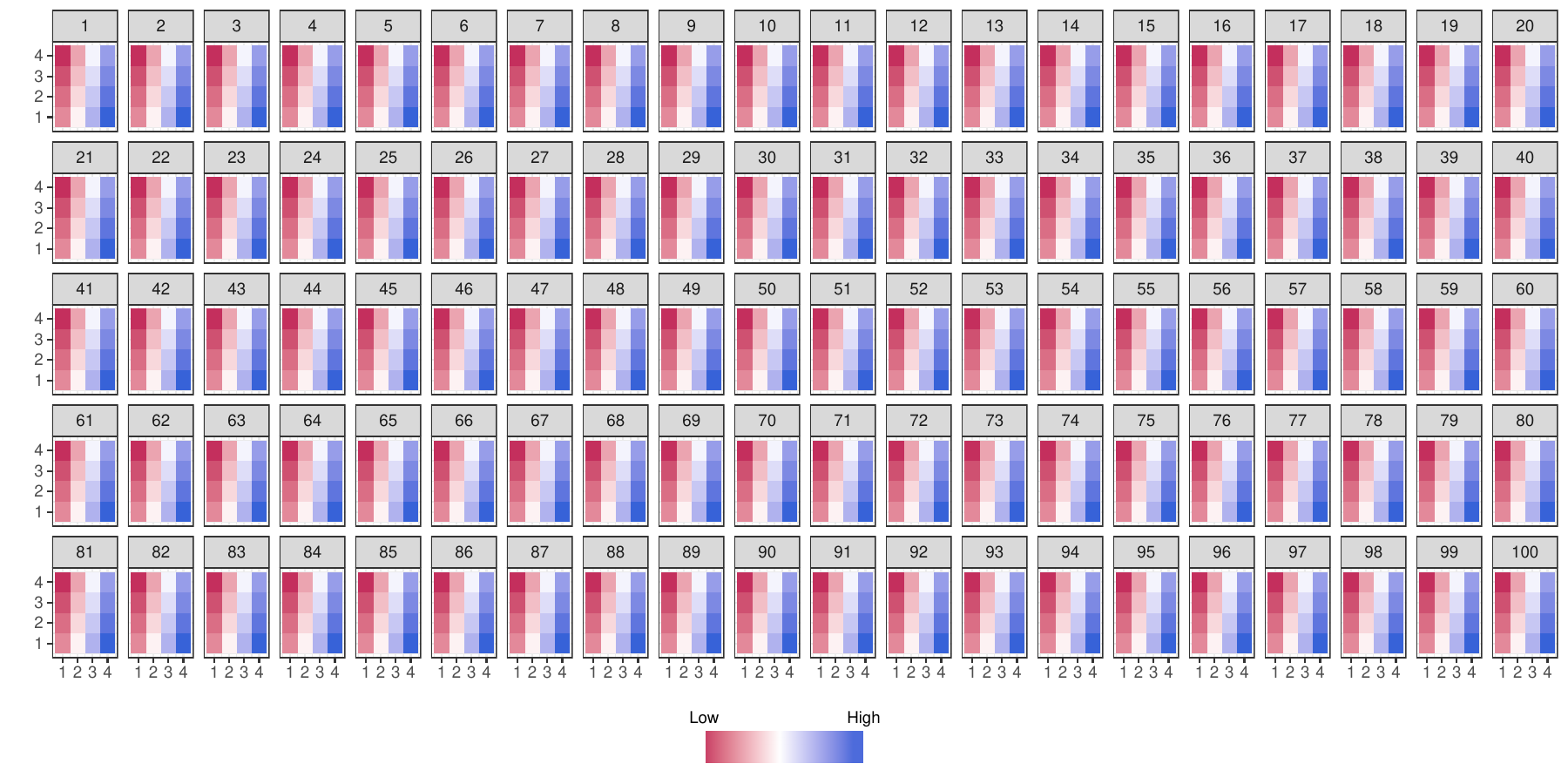} 		
\captionof{figure}{Transplantation priorities of waiting time and health state combinations with the original LAS \label{LAS_1yr}}
\end{figure}

\begin{figure}[H]
\includegraphics[width=0.98 \linewidth]{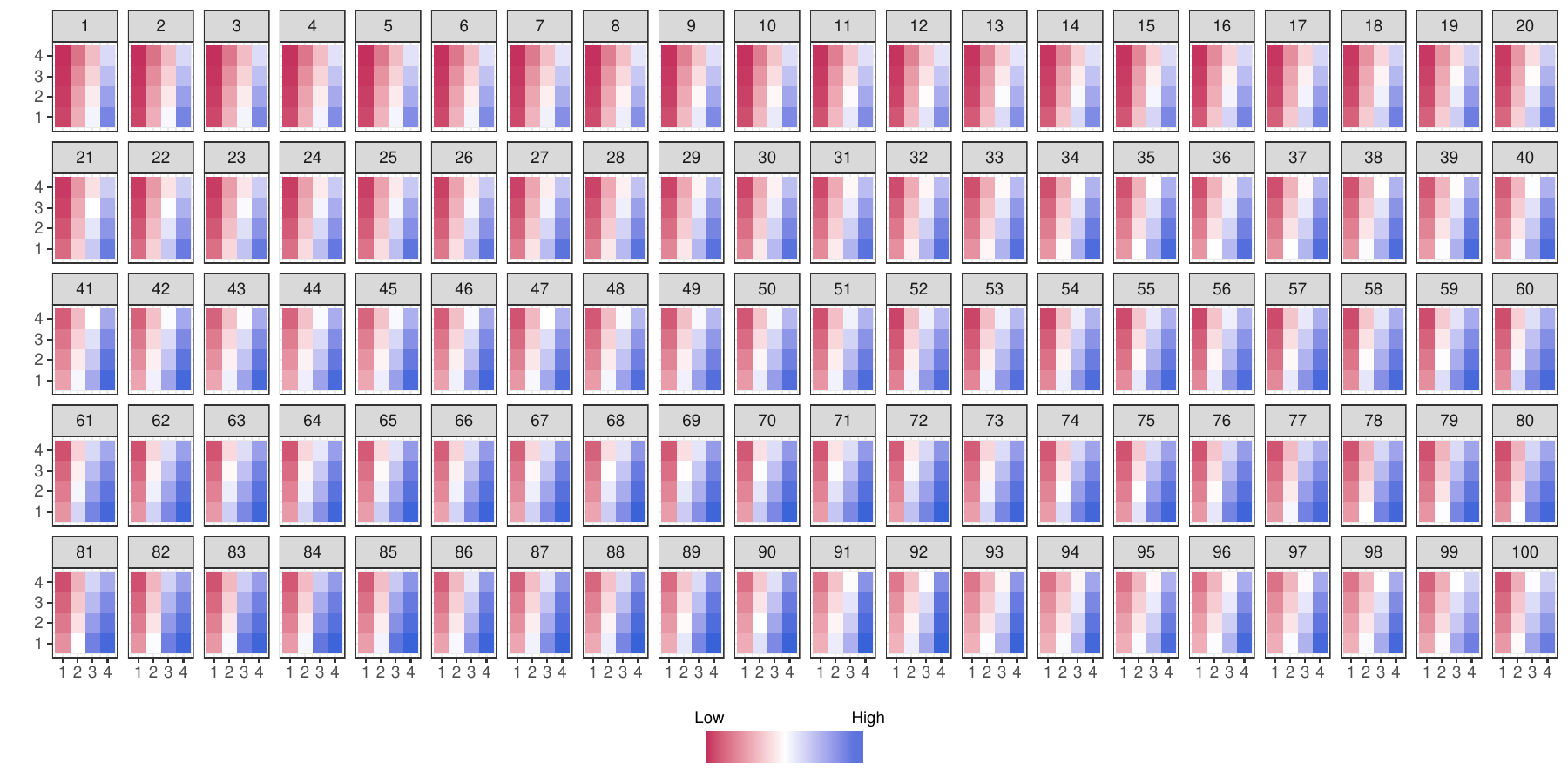} 		
\captionof{figure}{Transplantation priorities of waiting time and health state combinations with the refined LAS without one-year constraint and without the waiting-time-invariance assumption\label{LAS2}}
\end{figure}

See Figures \ref{Optimal}, \ref{LAS_1yr} and \ref{LAS2} below for a comparison of priority rankings with the proposed allocation strategy and the two LAS-type methods. In each of the graphs, numbers $1-100$ floating on the gray background indicate the discretized waiting time periods of $30$ days. Each mosaic pattern under the waiting time title contains the $4 \times 4$ state space of $(S_{wl}, S_{\mu})$, where the $x$-axis indicates the states of $S_{wl}$ and the $y$-axis indicates the states of $S_{\mu}$. Larger index of the states corresponds to shorter life.
Therefore, each small rectangle in the graph shows the priority of a combination of waiting time and health states, where blue indicates higher priority for transplantation, while red indicates lower priority. For example, the rectangle on the very top-left of Figure \ref{Optimal} is in deep red, demonstrating that with the proposed allocation strategy, patients that are in the first $30$-day waiting time period and with $(S_{wl}, S_{\mu}) = (1, 4)$ have very low priority of transplantation. This is reasonable, since $S_{wl} = 1$ indicates a prediction of long in-waiting-list residual life while $S_{\mu} = 4$ predicts a short post-transplant residual life, and thus low benefit of transplantation, especially given the fact that these patients are new to the waiting list and are less urgent in receiving transplants.

\subsection{Comparison of Policy Performances via Simulation Results}

Finally, we simulate waiting lists with parameters estimated from the UNOS data. 
Specifically, patient and organ arrivals are simulated according to independent homogeneous Poisson processes. The organ arrival rate $\rho$ and the patient arrival rate $\tau$ are estimated with average numbers of patient arrivals and organ arrivals per $ 30 $-day period after year 2006: $\hat{\rho}=104$, $\hat{\tau}=173$.
At the end of each waiting time period, counterfactual transitions in patients' health states are simulated according to the estimated transition probabilities. Using priority rankings of the proposed allocation strategy and the two LAS methods, organs, whenever available, are allocated to patients with the highest ranked combinations of waiting times and states in the current waiting list. Patients who are not selected for transplants remain on the waiting list with possibilities of future counterfactual transitions and transplantations. At the end of the simulation, the total life, life in the waiting list and post-transplant life, averaged over all patients ever entered the waiting list, are counted in days for each of the allocation strategies.

Table \ref{table: results} reports the means and standard deviations of averaged life outcomes of $200$ independent simulations. In each of the $200$ simulation runs, a waiting list containing $1605$ patients was generated and $973$ organs were allocated to the patients with each allocation strategy.
In addition to the above mentioned strategies, we consider a random allocation strategy in which patients are chosen for transplantation randomly.
In each simulation run, the generated waiting list is copied four times to apply each of the allocation strategies in question.  
The averaged total life under each allocation strategy in Table \ref{table: results} is also compared to the model upper bound of $1881$ days.

% results of average life (detailed)

\begin{table}[H]
\begin{tabular}{|p{2cm}|p{3cm}|p{3cm}|p{3cm}|}
\hline
& Average Total Life (SD) & Average Life in Waiting List (SD) & Average Life Post-Transplant (SD)  \\ \hline
Proposed & 1839 (19) & 489 (21) & 1349 (20)  \\ \hline
LAS & 1708 (20) & 554 (20) & 1155 (15) \\ \hline
Refined LAS & 1723 (21) & 561 (21) & 1162 (16) \\ \hline
Randomized & 1612 (26) & 469 (21) & 1143 (30) \\ \hline
\end{tabular}
\captionof{table}{Means and standard deviations (SD) of averaged life outcomes (in days) with different allocation strategies over $200$ independent simulations. } \label{table: results}
\end{table}

By applying the proposed allocation strategy, the averaged total life increased by $7.7\%$ comparing to the result with the LAS. The gain in total life came from a much improved post-transplant life: it increased by $16.8\%$ comparing to the post-transplant life under the LAS. Meanwhile, the average time in waiting list for patients (including those who never received transplantation and the organ recipients) was shortened by $11.7\%$. 
Applying the refined LAS without the one-year constraint in the LAS calculation and without the assumption that the waiting-list survival is invariant to the current waiting time slightly improves the averaged total life. As expected, the purely random allocation had a much worse performance than the proposed strategy and the LAS methods.

\begin{figure}[H] 
	\centering
	\includegraphics[width=1 \textwidth]{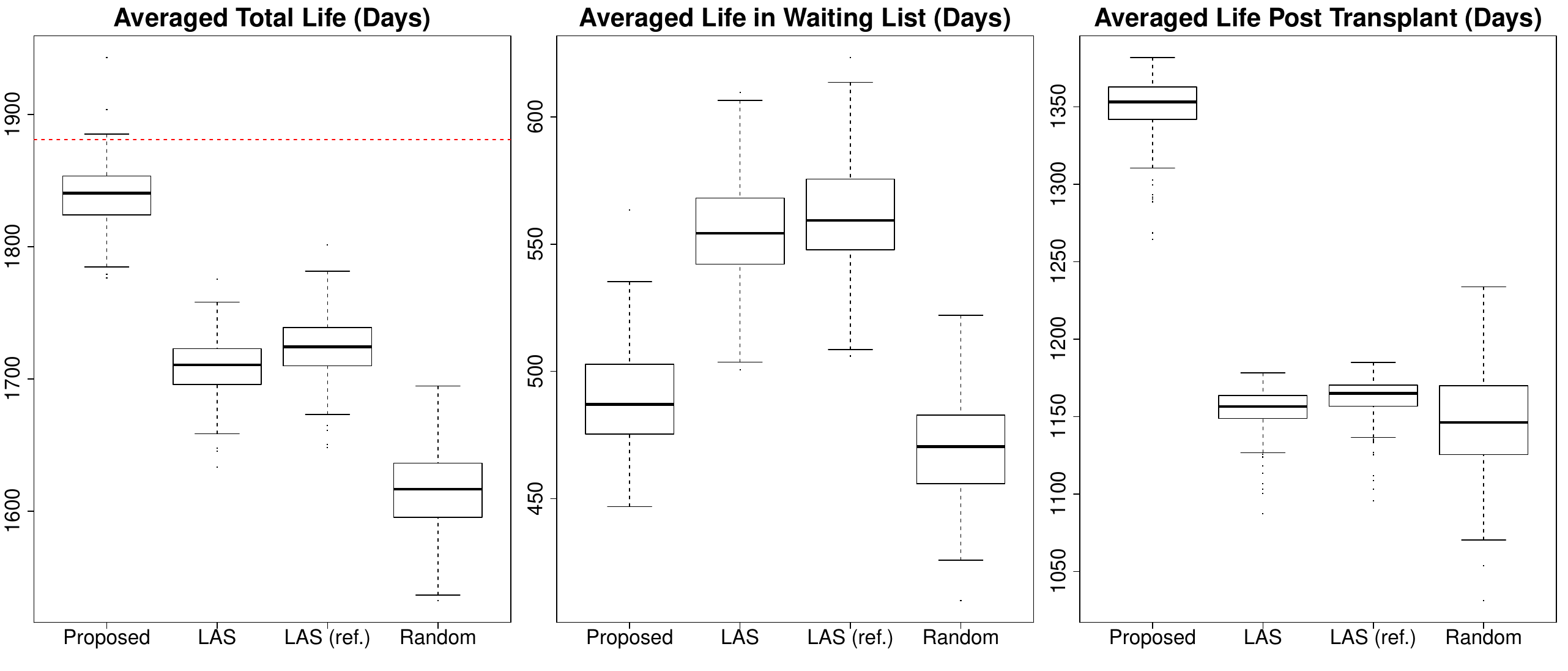}
	\captionof{figure}{Average Life with Different Allocation Strategies} \label{graph: results}
\end{figure}

Figure \ref{graph: results} shows the box plots of the averaged life outcomes with each allocation strategy. From left to right the averaged total life with the model upper bound of $1881$ (red dashed line), the averaged life in waiting list, and the averaged life post transplant are displayed.
An interesting observation is that while the randomized strategy resulted in a larger variance, especially in post-transplant life, which is expected given that organ recipients were selected randomly, the mean post-transplant life of patients under the random strategy was not substantially different from those under the LAS methods. This observation might indicate the current LAS system still has much space to improve, especially in extending patients' post-transplant life.

%---------------------------------------------------------------------------------
% Detailed in-waiting list life outcomes

\begin{figure}[H] 
\centering
\includegraphics[width=1\textwidth]{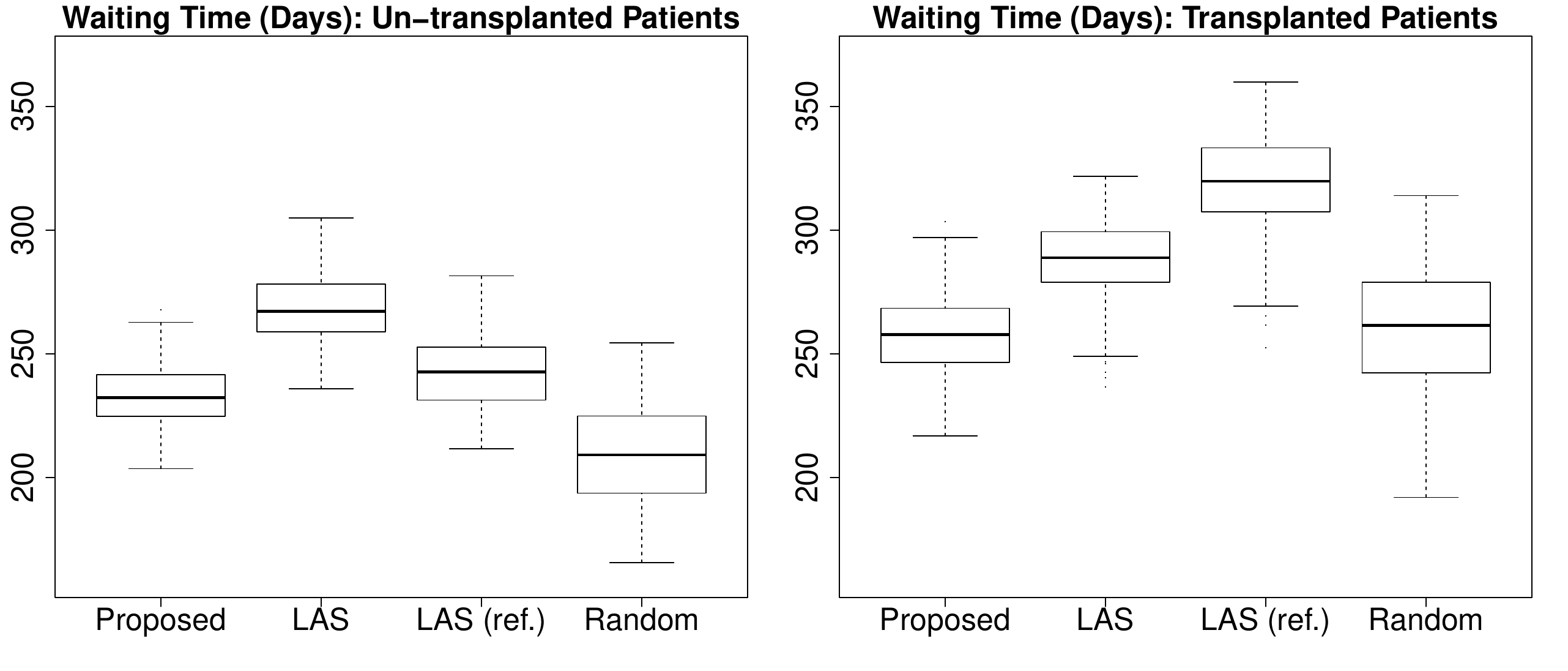}
\captionof{figure}{Average Waiting Time with Different Allocation Strategies. Left: patients died in the waiting list and never received transplants. Right: transplanted patients.} \label{graph: results_wl}
\end{figure}

Regarding patients' average life in the waiting list with each allocation strategy, Table \ref{table: results_wl} and Figure \ref{graph: results_wl} show detailed average waiting time for organ recipients and patients who never received transplantation. 

For patients who died while waiting for transplantation, the original LAS with one-year constraint led to the longest in-waiting-list life, possibly due to its emphasis on in-waiting-list survival. The LAS gives higher priority of transplantation to patients with large probability of dying within one year. Without the one-year constraint, the refined LAS allocated less organs to those who are likely to die in one year without transplant and hence the shorter average in-waiting-list life. 
Overall, the LAS methods emphasize more on the in-waiting-list survival over the post-transplant survival.

On the other hand, the proposed strategy aims to optimize patients' total life and thus patients with larger expected life extension with transplantation were given higher priority. Since some of the patients who were likely to die in one year without transplantation were also expected to have short post-transplant life, they were not selected as organ recipients, which resulted in a shorter average waiting list life. 

For patients who received transplantation, the proposed method shortened the average waiting time for an organ comparing to the LAS methods, which might be considered to be an advantage by the clinicians and patients.
The waiting time was shorter with the original LAS comparing to the refined LAS, which was possibly due to allocations of organs at early waiting time to patients who were otherwise going to die soon. 

\begin{table}[H]
\begin{tabular}{|p{2cm}|p{3.5cm}|p{3.5cm}|}
\hline
 & Average waiting time (un-transplanted) (SD)& Average waiting time (transplanted) (SD)\\ \hline
Proposed & 234 (14) & 256 (18) \\ \hline
LAS & 267 (14) & 286 (17) \\ \hline
Refined LAS & 242 (15) & 318 (20) \\ \hline
Randomized & 213 (17) & 257 (22) \\ \hline
\end{tabular} 
\captionof{table}{Comparison of  Average Waiting Time with Different Allocation Strategies} \label{table: results_wl}
\end{table}

\section{Numerical Results on Multiple Organ Type Matching} \label{sec: multiple numeric}

Obtaining substantive results in the multiple organ type scenario would require a comprehensive study of organ properties from real transplantation data. 
Here we demonstrate the proposed allocation strategy for multiple organ types with a synthetic data example. 

Two organ types are generated in the example.
The post-transplant residual life with type-$1$-organ transplantation is larger for all $(i, s)$ combinations comparing to with type-$2$-organ transplantation. 
However, for some $(i, s)$ combinations, the difference in the benefit between the two organ types is less significant, modeling the situations where patients in certain states at certain stage of the disease may have a better compatibility with type $2$ organs.
This example is generated to resemble the phenomenon that patients generally have a better survival if transplanted with non-smoking-donor organs, but recipients with certain characteristics may suffer less from the reduction in benefit when transplanted with smoking-donor organs.

Following the three-stage algorithm in Section \ref{sec: multi organ strategy} motivated by the monotonicity properties Propositions \ref{thm: weak monotone 1} and \ref{thm: weak monotone 2}, the partition of the space of $(i, s)$ combinations into disjoint subsets by their designated organ types is obtained, and complete rankings of priorities for each organ type are calculated.
All results are calculated with $\sum_{w=1}^2 \rho^{(w)} / \tau = 104/173$ and $\rho^{(1)}/\rho^{(2)} = 1$. 

\begin{figure}[H]
	\includegraphics[width=1\linewidth]{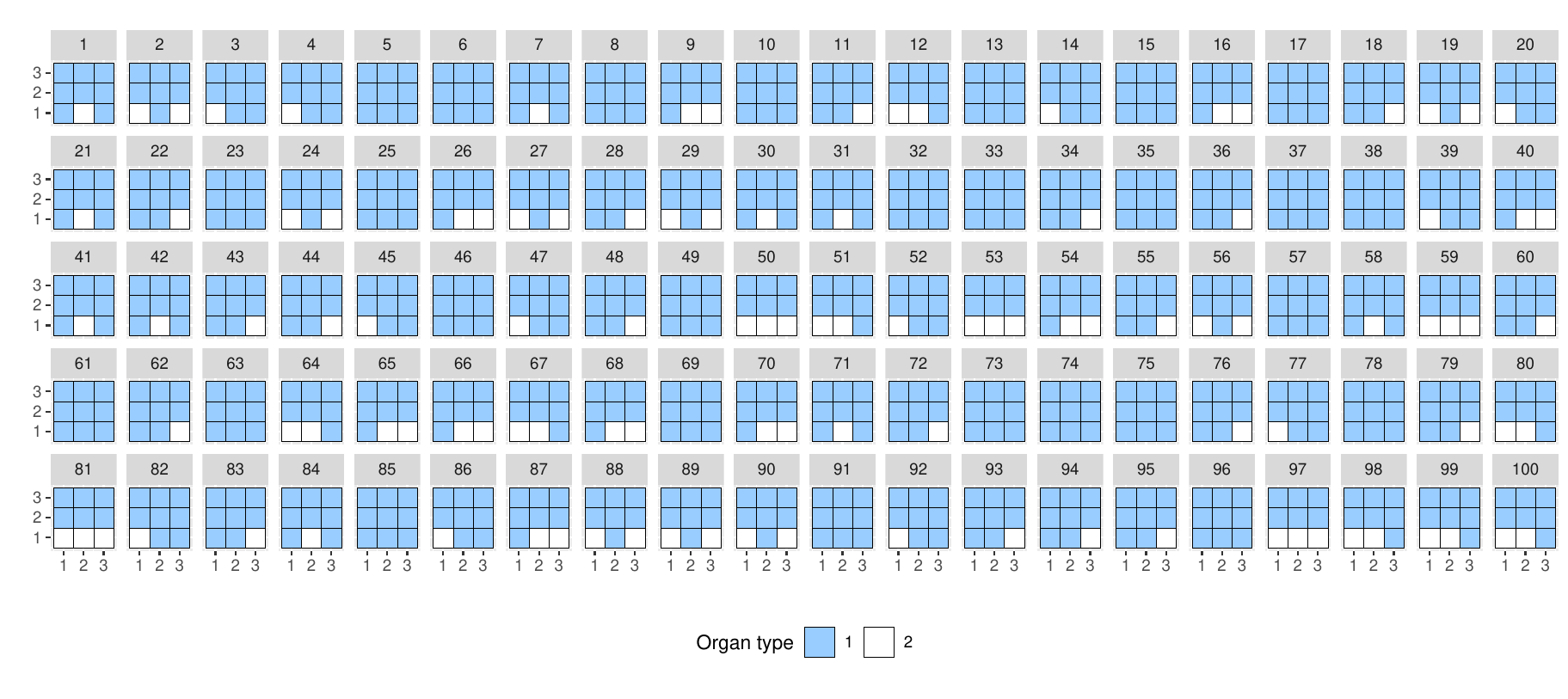} 		
	\captionof{figure}{The partition of the space of health state and waiting time combinations according to their designated organ types \label{partition}} 
\end{figure}

\begin{figure}[H]
	\includegraphics[width=1\linewidth]{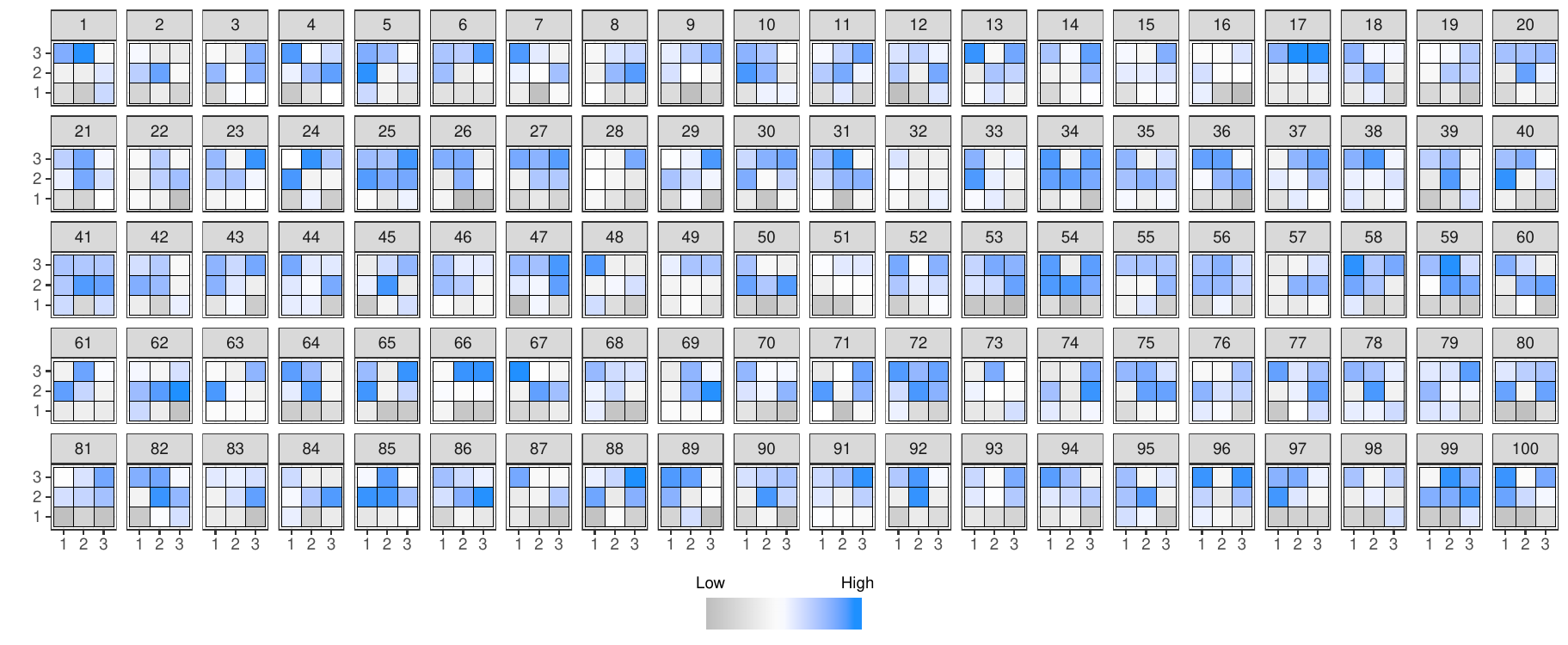} 		
	\captionof{figure}{Difference between post-transplant residual life with type $1$ and type $2$ organs for each $(i, s)$ combination. Dark gray color indicates smaller difference while blue color indicates larger difference in residual life. \label{Ucompare}} 
\end{figure}

\begin{figure}[H]
	\includegraphics[width=1\linewidth]{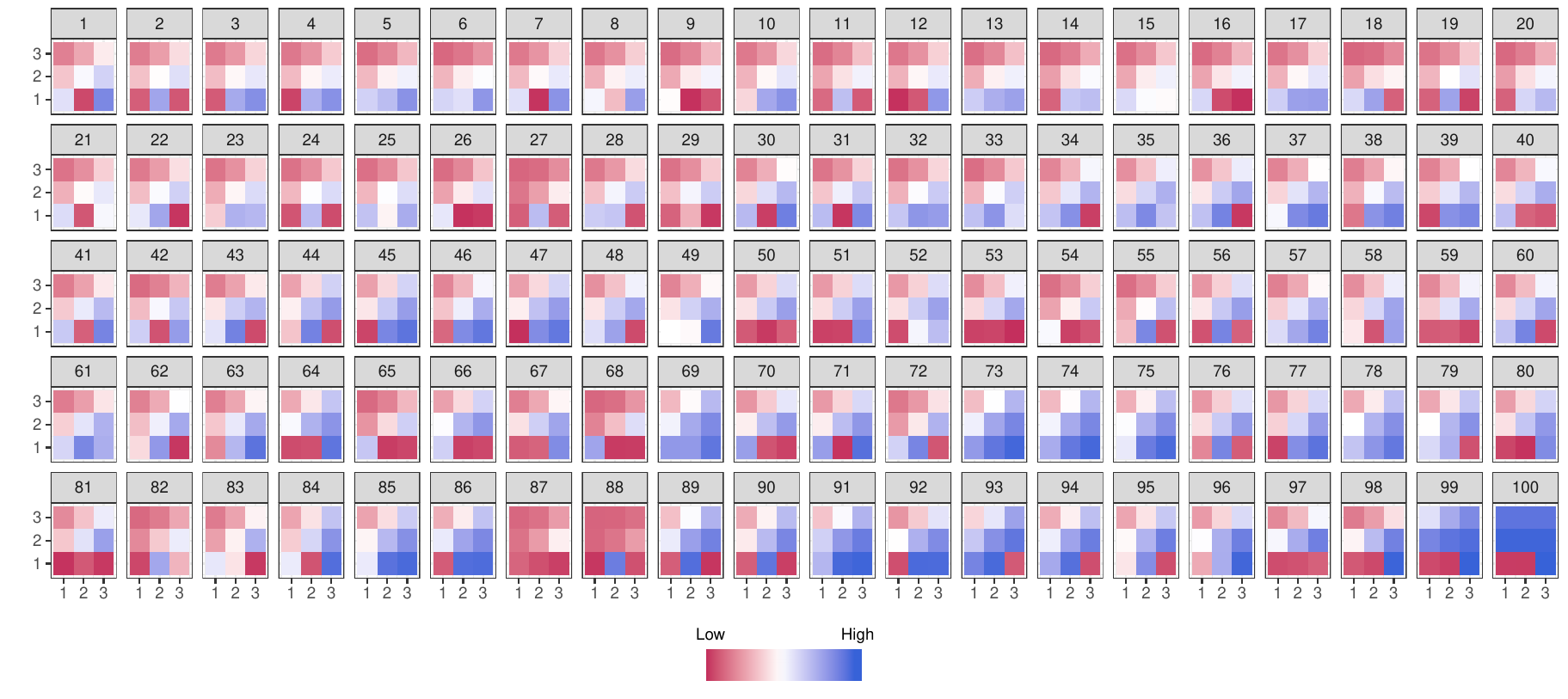} 		
	\captionof{figure}{Transplantation priorities of waiting time and health state combinations with the proposed strategy for type 1 organs  \label{rank_organ_type_1}} 
\end{figure}

\begin{figure}[H]
	\includegraphics[width=1\linewidth]{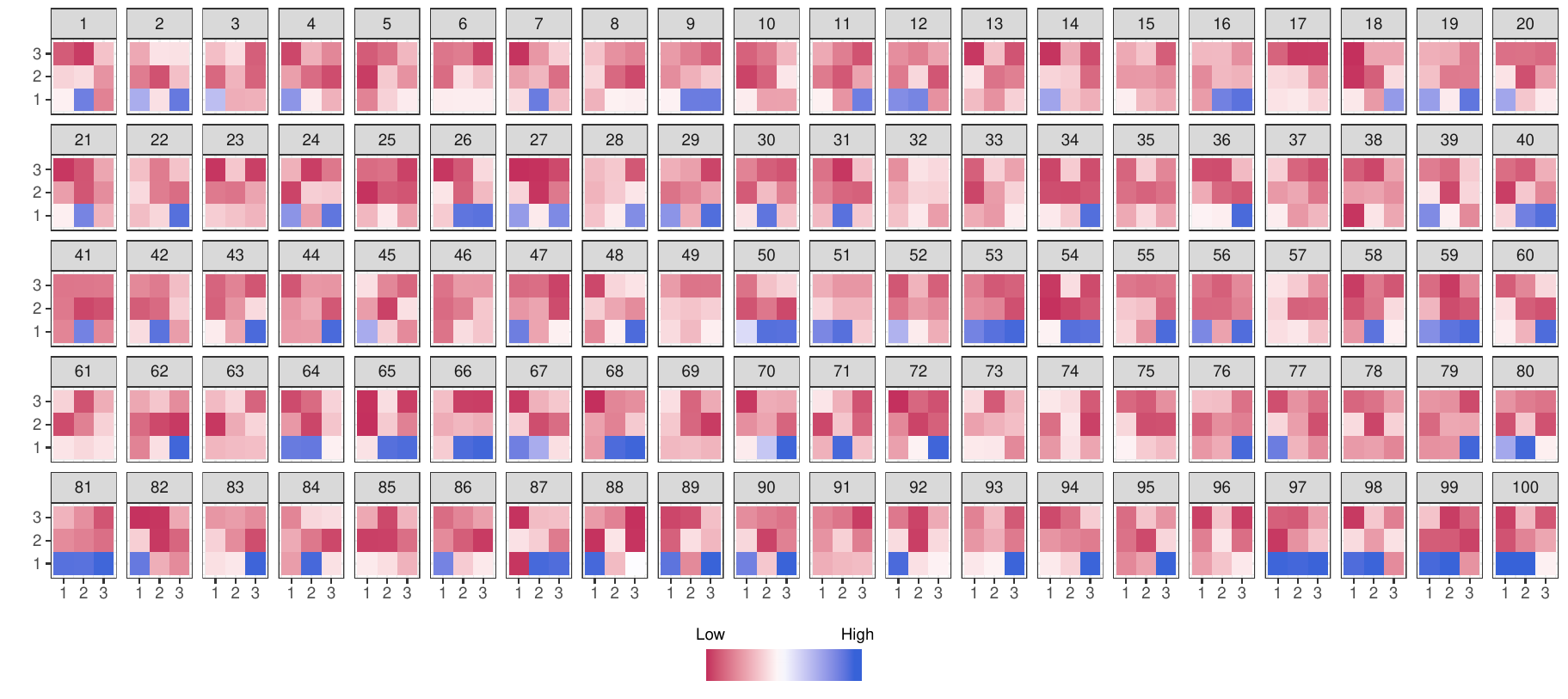} 		
	\captionof{figure}{Transplantation priorities of waiting time and health state combinations with the proposed strategy for type 2 organs  \label{rank_organ_type_2}} 
\end{figure}

Results are visualized in Figures \ref{partition} - \ref{rank_organ_type_2}.
In each of the graphs, numbers $1-100$ floating on the gray background indicate the discretized waiting time periods of $30$ days. 
Each mosaic pattern under the waiting time title contains the $3 \times 3$ state space of $(S_{wl}, S_{\mu})$, where the $x$-axis indicates the states of $S_{wl}$ and the $y$-axis indicates the states of $S_{\mu}$.

Figure \ref{partition} shows the partition of $(i, s)$ combinations by designated organ type. 
The majority of $(i,s)$ combinations have color blue, indicating type 1 (non-smoking-donor) organ is the designated organ type. 
However, for some $(i,s)$ combinations the designated organ type is type $2$ (smoking-donor) organ. 
This partition comports with the differences between post-transplant residual life with type $1$ and type $2$ organs visualized in Figure \ref{Ucompare}. 
In particular, positions of the white rectangles in Figure \ref{partition} match those of the dark gray rectangles in Figure \ref{Ucompare}, which are $(i,s)$ combinations that suffer less from the reduction in post transplant residual life with type $2$ organs.

Figures \ref{rank_organ_type_1} and \ref{rank_organ_type_2} visualize the priority rankings of $(i,s)$ combinations for transplantation with type $1$ and type $2$ organs respectively. 
In Figure \ref{rank_organ_type_1} it is expected that the priority is lowest whenever the color is white in the partition in Figure \ref{partition}, 
as those $(i,s)$ combinations have designated organ type 2 and will only be considered for type-$1$-organ transplantation if no patients in $(i, s)$ colored blue, indicating type 1 as designated organ type, are available in the current waiting list when a type 1 organ arrives. 
Among those $(i,s)$ combinations with designated organ type 1, the priority ranking of transplantation generally resembles the ranking in the single-organ case: $(i,s)$ with longer estimated post transplant residual life and shorter in-waiting-list residual life, such as the bottom-right rectangle in each mosaic, generally has higher priority of transplantation.
Similar phenomena can be observed in the priority ranking of type 2 organ transplantation in Figure \ref{rank_organ_type_2}.

However, the priority rankings in the multiple organ type scenario do not always follow those in the single organ type case. ``Flips'' of the priorities are observed: some $(i, s)$ combinations with high priorities in the single organ type case now have lower priorities comparing to other combinations. Examples of such flips include the bottom row of mosaic No. 83 and mosaics No. 87 and 88 in Figure \ref{rank_organ_type_1}. 
This phenomenon of flips echos our discussion on the monotonicity in Section \ref{sec: multi organ strategy}: the priority ranking for one organ type is affected by the relative availability of all organ types represented by $\mathbf{c}$ and the strong monotonicity of priorities usually does not hold when $\mathbf{c}$ decreases freely. 

Overall, the above results demonstrate that multiple organ type allocation and matching is not a trivial extension of the single organ type problem. 
The ranking of $(i,s)$ combinations for each organ type depends on the partition of $(i,s)$ combinations by their designated organ types. 
Moreover, the partition and priority rankings are determined by the complex interplay between transitions in health states, the residual life of each state at each waiting time, and the relative availability of different organ types.

\section{Concluding Remarks}

We have presented an approach to modeling the lung transplantation waiting list and comparing allocation rules.
Here patient and organ arrivals are modeled as independent homogeneous Poisson processes, and counterfactual patient health status trajectories absent transplantation are modeled as independent and identically distributed inhomogeneous Markov processes. Patients' expected post-transplantation residual life depends on both the health state at the time of transplantation and the waiting time at transplantation.  
The model setting here is capable of capturing the randomness in patient and organ arrivals and the complex dynamics of patients' health characteristics and their effects on life outcomes.  
In practice, the researchers and policy makers can always expand the state space of patient health state so that trajectories of the transitions of health states approximately follow the Markov property. For example, previous medical records of patients can be included if they are believed to contain important information for predicting future transitions. 

Allocation rules are modeled as index sequences of transplanted patients. 
Only fair allocation rules are considered in the comparison. Under fair allocation rules, the choice of patients for transplantations are decided by patients' health states and waiting times at the time of organ arrival and a random variable that is conditionally independent of patients' past and future states and survivals. Allocation probabilities are also required to be invariant to patient index and calendar time. This definition of fairness has two implications: first, unrealistic rules that can predict patients' future states and survivals are not considered; second, given patients' health states and waiting times, the allocation decisions are independent of other factors. Therefore the definition of fairness here implicitly addresses the equity issue in organ allocations. 

It is shown each fair allocation rule has a corresponding limiting transplantation rate. Under a fair allocation rule, the average rate of transplantation to patients in any state and waiting time converges to the corresponding transplantation rate as calendar time increases. 
The limiting transplantation rate satisfies constraints that reflect the scarcity of organs. The main constraint is that the average proportion of transplanted patients in the limit is bounded by the ratio of organ arrivals to that of patient arrivals.
The limiting average total life (or equivalently, the limiting average life gain from transplantation), represented in terms of the transplantation rate, is used as the standard in comparing fair allocation rules. 

The optimal transplantation rate subject to the constraints is characterized recursively with the Hamilton-Jacobi-Bellman equations. Then a fair allocation strategy is developed based on the form and monotonicity properties of the optimal transplantation rate. 
The allocation strategy is to use the penalty parameter $c$ associated with the constraint on the average proportion of transplanted patients as an index to prioritize patients' states and waiting times. 

The index $c$ is partially related to the Gittins Index (\cite{Gittins:1989ui}).
\cite{Bertsimas:2000iz} considered a setting that is close to the organ transplantation problem, in which a fixed number of resources are to be allocated to a fixed number of subjects that are in constant transitions of states in discrete time. 
Allocating resources using Gittins Index leads to optimal or asymptotically optimal objectives in problems where subjects remain static if not selected for allocations (\cite{Weber:1992ue}, \cite{Whittle:1980tz}, \cite{Whittle:1988un}) and in settings where the rates of patient and organ arrivals tend to infinity. In restless bandits problems in which all subjects are constantly in transitions, including the problem studied here, though there are sufficient conditions for Gittins Index to be optimal (\cite{Bertsimas:1996ke}, \cite{NinoMora:2001vw}), the optimality is not guaranteed in general. 

Existing restless bandits models do not cover all complexities of the lung transplantation allocation problem. These complexities include the random arrival of patients and organs to the waiting list, random exits of patients due to death and transplantation, constraints on the allocation when an organ becomes available due to limited availability of patients in allocable states, and fairness considerations. 

In our paper we model the lung transplant problem accounting for these complexities and study asymptotic properties of the waiting list. Then we provide a heavy traffic solution to optimize patients' long-term average life.
The heavy traffic optimal solution given in Section \ref{sect: optimization} can be seen as a much generalized version of the Gittins Index: instead of having a discrete number of jobs in the restless bandits problem, we have a continuum of ``jobs'' (expected occupancy of states at waiting times) and the resources are allocated to the continuum of jobs under constraints that are also continuous. As discussed in Section \ref{sect: strategy}, the heavy traffic optimal solution cannot be directly applied to the allocation of organs in reality, and we propose an allocation strategy motivated by the heavy traffic solution that can lead to a long-term average life that is close to the heavy traffic upper bound. 

In short, part of our development is tangentially related to the restless bandits problem and the Gittins Index. The scope of the lung transplantation problem itself, however, is beyond the idealized restless bandits model due to its complexities.

Simulation studies show it may be possible to improve the Lung Allocation Score (LAS) currently used by UNOS and increase the average total life by as much as 7.7\%. Results provided here are provisional and a deeper understanding of the lung allocation procedure and the optimal allocation strategy requires further effort. As discussed previously, there may be a gap between the objective using the proposed strategy and the practical upper bound of the objective. The gap may be a result of the lack of patients in optimal states and waiting times when organs are available, which may stem from the fact the constraints in the optimization problem do not cover all practical confinements on the allocations. 

In future studies we will extend the donor-recipient matching results developed here and conduct comprehensive substantive studies using not only patients but also donor records from UNOS.
Issues related to cross-region transplantation, or other practical aspects in lung transplantation may also require further constraints to be imposed in formulating the optimal transplantation rate and developing allocation strategies.

\section*{Acknowledgement}
The authors would like to thank Eric Peterson for his assistance with the UNOS data and Marco Scarcini for discussions on this problem. We would also like to thank the AE and the reviewers for their invaluable comments and suggestions in the process of improving this paper.

% AOS,AOAS: If there are supplements please fill:
%\begin{supplement}[id=suppA]
%  \sname{Supplement A}
%  \stitle{Title}
%  \slink[doi]{10.1214/00-AOASXXXXSUPP}
%  \sdatatype{.pdf}" 
%  \sdescription{Some text}
%\end{supplement}

\newpage

\section*{Appendix} 

\subsection{Coefficient Estimates for In-Waiting-List and Post-Transplant Survivals}

\begin{center}	
	\LTcapwidth=\textwidth
	\begin{longtable}{|p{5cm}|p{1cm}|p{2.5cm}|p{3cm}|} 
		\caption{Coefficient Estimates for In-Waiting-List Hazard for Death \label{t1}} \\
		\hline 
		Covariate & Group & Coefficient & $ p $-value\footnote{Significance level: *** $ p $-value $ \le 0.001 $, ** $ 0.001 < $ $ p $-value $ \le 0.01 $,  * $ 0.01 < $ $ p $-value $ \le 0.05 $,  . $ 0.05 < $ $ p $-value $ \le 0.1 $. }\\
		\hline 
		\hline 
		Age (year) & 3 & 0.014 & $ < 2\times 10^{-16} $ ({*}{*}{*})\\
		\hline 
		Body mass index (BMI; $kg/m^{2}$) & 1 & 0.116 $ \times $ (20 - BMI) for BMI less than 20 $kg/m^{2}$ &  $< 2\times 10^{-16} $ ({*}{*}{*}) \\
		\hline 
		Ventilation status if candidate is hospitalized & 1 & -0.444 if continuous mechanical ventilation needed &  $6.7\times 10^{-14} $ ({*}{*}{*})  \\
		\hline 		
		Creatinine (serum, mg/dL) & 1 & 0.220 if at least 18 years of age & $< 2\times 10^{-16} $ ({*}{*}{*})\\
		\hline 
		Diabetes & 1 & 0.173 & $< 2\times 10^{-16} $ ({*}{*}{*})\\
		\hline 
		Diagnosis Group A & 3 & 0 & NA\\
		\hline 
		Diagnosis Group B & 3 & 0.794 & $< 2\times 10^{-16} $ ({*}{*}{*}) \\
		\hline 
		Diagnosis Group C & 3 & 1.126 & $< 2\times 10^{-16} $ ({*}{*}{*}) \\
		\hline 
		Diagnosis Group D & 3 & 0.163 & $ 3.3\times 10^{-4} $ ({*}{*}{*})\\
		\hline
		 Detailed diagnosis: Bronchiectasis (Diagnosis Group A only) & 3 & 0.182 & $ 1.4\times 10^{-3} $ ({*}{*}{*}) \\
		\hline  
		 Eisenmenger's syndrome (Diagnosis Group B only) & 3 & -1.04 & $< 2\times 10^{-16} $ ({*}{*}{*})\\
		\hline 
		 Lymphangioleiomyomatosis (Diagnosis Group A only) & 3 & -0.961 & $ 1.7\times 10^{-13} $ ({*}{*}{*})\\
		\hline 
		 Obliterative bronchiolitis (Diagnosis Group D only) & 3 & -0.416 & $ 8.3\times 10^{-5} $ ({*}{*}{*})\\
		\hline 
		 Pulmonary fibrosis, not idiopathic (Diagnosis Group D only) & 3 & 0.014 & 0.70\\
		\hline 
		 Sarcoidosis with PA mean pressure greater than 30 mm Hg (Diagnosis Group D only) & 3 &  -0.44 & $< 2\times 10^{-16} $ ({*}{*}{*}) \\
		\hline 
		 Sarcoidosis with PA mean pressure of 30 mm Hg or less (Diagnosis Group A only) & 3 & 0.613 & $< 2\times 10^{-16} $ ({*}{*}{*}) \\
		\hline 
		Forced vital capacity (FVC) & 2 & 0.188 $ \times $ (80 - FVC)/10 if FVC is less than 80\% for Diagnosis Group D & $< 2\times 10^{-16} $ ({*}{*}{*})\\
		\hline 
		Functional Status & 1 & -0.287 if no assistance needed with activities of daily living & $< 2\times 10^{-16} $ ({*}{*}{*})\\
		\hline 
		Oxygen needed to maintain adequate oxygen saturation (80\% or greater)
		at rest (L/min) & 2 & 0.111 for Group B, 0.108 for Groups A, C, and D & $< 2\times 10^{-16} $ ({*}{*}{*}) \\
		\hline 
		PCO2 (mm Hg): current & 1 & 0.222 if PCO2 is at least 40 mm Hg & $< 2\times 10^{-16} $ ({*}{*}{*})\\
		\hline 
		PCO2 increase of at least 15\% & 1 & -0.232 if PCO2 increase is at least 15\% & $ 9.4\times 10^{-15} $ ({*}{*}{*})\\
		\hline 
		Pulmonary artery (PA) systolic pressure (10 mm Hg) at rest, prior
		to any exercise & 2 & 0.003 for Group A if the PA systolic pressure is greater than 40 mm
		Hg, 0.016 for Groups B, C, and D &  $< 2\times 10^{-16} $ ({*}{*}{*}),  $ 9.3\times 10^{-3} $ ({*}{*}{*})\\
		\hline 
		Six minute walk distance (feet) obtained while the candidate is receiving
		supplemental oxygen required to maintain an oxygen saturation of 88\%
		or greater at rest. Increase in supplemental oxygen during this test
		is at the discretion of the center performing the test. & 1 &  -0.075 $ \times $ Six-minute-walk distance/100 & $< 2\times 10^{-16} $ ({*}{*}{*})\\
		\hline 
	\end{longtable}
\end{center}

\begin{center}	
	\LTcapwidth=\textwidth
	\begin{longtable}{|p{5cm}|p{1cm}|p{2.5cm}|p{3cm}|}
		\caption{Coefficient Estimates for Post-Transplantation Hazard for Death \label{t2}} \\
		\hline 
		Covariate & Group & Coefficient & $ p $-value\footnote{Significance level: *** $ p $-value $ \le 0.001 $, ** $ 0.001 < $ $ p $-value $ \le 0.01 $,  * $ 0.01 < $ $ p $-value $ \le 0.05 $,  . $ 0.05 < $ $ p $-value $ \le 0.1 $. }\\
		\hline 
		\hline 
		Age (years) & 4 &  $4.4\times 10^{-3}  \times $ (age - 45) if greater than 45 years of age & $  2.1\times 10^{-5} $ ({*}{*}{*})\\
		\hline 
		Creatinine (serum) at transplant (mg/dL) & 1 & 0.177 if candidate is at least 18 years old  & 0.003 ({*}{*})\\
		\hline 
		Creatinine increase of at least 150\% & 1 & 0.570 if increase in creatinine is at least 150\%, and the higher value determining this increase is at least 1 mg/dL  & 0.422 \\
		\hline 
		Ventilation status if candidate is hospitalized & 1 & -0.05 if continuous mechanical ventilation needed & 0.710\\
		\hline 
		Diagnosis Group A & 4 & 0 & NA\\
		\hline 
		Diagnosis Group B & 4 & 0.263 & 0.017 ({*})\\
		\hline 
		Diagnosis Group C & 4 & 0.268 & 0.004 ({*}{*})\\
		\hline 
		Diagnosis Group D & 4 & 0.171 & 0.009 ({*}{*})\\
		\hline 
		 Detailed diagnosis: Bronchiectasis (Diagnosis Group A only) & 4 &0.191 & 0.168\\
		\hline 
		 Eisenmenger's syndrome (Diagnosis Group B only) & 4 &  0.745 & 0.297\\
		\hline 
		 Lymphangioleiomyomatosis (Diagnosis Group A only) & 4 & -0.625  & 0.052 (.)\\
		\hline 
		 Obliterative bronchiolitis (not-retransplant, Diagnosis Group D only) & 4 & 0.035 & 0.866\\
		\hline 
		 Pulmonary fibrosis, not idiopathic (Diagnosis Group D only) & 4 & -0.150 & 0.077 (.)\\
		\hline 
		 Sarcoidosis with PA mean pressure greater than 30 mm Hg (Diagnosis Group D only) & 4 & -0.230 & 0.078 ({.})\\
		\hline 
		 Sarcoidosis with PA mean pressure of 30 mm Hg or less (Diagnosis Group A only) & 4 & -0.043 & 0.801\\
		\hline 	
		Oxygen needed to maintain adequate oxygen saturation (80\% or greater)	at rest (L/min) & 4 & $ 6.6 \times 10^{-3} $ for Group A 		
		$ 1.1\times 10^{-3} $ for Groups B, C, and D & 0.590, 0.845\\
		\hline 
		Functional Status & 1 & -0.206 if no assistance needed with activities of daily living & 0.001 (***)\\
		\hline 
		Six minute walk distance (feet) obtained while the candidate is receiving
		supplemental oxygen required to maintain an oxygen saturation of 88\%
		or greater at rest. Increase in supplemental oxygen during this test
		is at the discretion of the center performing the test. & 1 & $ 3.0\times 10^{-4} \times $  (1200-Six minute walk distance), 0 if six-minute-distance-walked
		is at least 1200 feet & $ 6.2\times 10^{-10} $ ({*}{*}{*})\\
		\hline 
	\end{longtable}	
\end{center}

\subsection{Proofs of Main Technical Results}
Here we outline the proofs of selected main results. For detailed proofs of all technical results, see \cite{Zou:2015ur}.

\subsubsection*{Proof of Theorem \ref{thm: soln}}

\begin{proof}

For any $i \in \mathcal{X}$, $\Psi_{i}$ can be decomposed into (c.f. for example \citet{Halmos:1974uy}), 
\[
\Psi_{i}=\Psi_{i}^{ac}+\Psi_{i}^{s},
\]
where $\Psi_{i}^{ac}$ is absolutely continuous with respect to the Lebesgue measure with density $\psi_{i}$ and $\Psi_{i}^{s}$
is singular to the Lebesgue measure with support $B \subset \mathbb{R}$. 

Define the value function 
\begin{equation} \label{eq: value fn}
V_{s-}(\boldsymbol{\uppi})=\sup_{\{\boldsymbol{\Psi}_{t}:s\leq t\leq T\}} \Big\{\int_{[s,T]} d\boldsymbol{\Psi}_{t}\cdot(\tilde{\boldsymbol{\upmu}}_t-c\cdot\mathbf{1}_n)\Big\},
\end{equation}
where $ d\boldsymbol{\Psi}_{t}=\boldsymbol{\uppi}_{t-}d\boldsymbol{Q}_{t}-d\boldsymbol{\uppi}_{t} $, in which $\boldsymbol{\uppi}_{s-}=\boldsymbol{\uppi}$, $\boldsymbol{\uppi}_{t}=\boldsymbol{\uppi} \cdot\prodi_{[s,t]}(\boldsymbol{I}+d\boldsymbol{Q})(\boldsymbol{I}-d\boldsymbol{\Lambda})$ for any $t\in[s,T]$. 

By the dynamic programming principle (see, for example, \citet{Fleming:2006tl}),
for all $s\in[0,T]$ and $\Delta s>0$, 
\begin{multline*}
V_{s-}(\boldsymbol{\uppi})=\sup_{\{\boldsymbol{\Psi}_{t}:s\le t<s+\Delta s\}}\Bigg\{ V_{(s+\Delta s)-}(\boldsymbol{\uppi}_{(s+\Delta s)-})\\
+\int_{[s,s+\Delta s)}(d\boldsymbol{\Psi}^{ac}(t)+d\boldsymbol{\Psi}^{s}(t))\cdot(\tilde{\boldsymbol{\upmu}}_t-c\cdot\mathbf{1}_n)\Bigg\}
\end{multline*}
which leads to 
\begin{multline*}
\sup_{\{\boldsymbol{\Psi}_{t}:s\le t<s+\Delta s\}}\Bigg\{\frac{V_{(s+\Delta s)-}(\boldsymbol{\uppi}_{(s+\Delta s)-})-V_{s-}(\boldsymbol{\uppi}_{s-})}{\Delta s}\\
+\frac{1}{\Delta s}\int_{[s,s+\Delta s)}(d\boldsymbol{\Psi}^{ac}(t)+d\boldsymbol{\Psi}^{s}(t))\cdot(\tilde{\boldsymbol{\upmu}}_t-c\cdot\mathbf{1}_n)\Bigg\}=0.
\end{multline*}

If $s\notin B$, $\int_{\{s\}}d\Psi_{i}^{s}(u)=0$ for all $i\in\mathcal{X}$, and the Radon-Nikodym derivative $\psi_{i}$ exists at $s$, in which case the value function is a solution to the Hamilton-Jacobi-Bellman equations
\begin{equation} \label{eq: HJB for V} 
\frac{\partial V}{\partial s}+\sup_{\{\boldsymbol{\Psi}_{t}:s\le t<s+\Delta s\}}\{\boldsymbol{\psi}_{s}\tilde{\boldsymbol{\upmu}}_s-c\cdot \boldsymbol{\psi}_{s}\mathbf{1}_n+\frac{\partial V}{\partial\boldsymbol{\uppi}}\cdot\frac{d\boldsymbol{\uppi}}{ds}\}=0,
\end{equation}
where 
\[
\frac{\partial V}{\partial\boldsymbol{\uppi}}=\int_{(s,T]}\Prodi_{(s,t)}(\boldsymbol{I}+d\boldsymbol{Q})(\boldsymbol{I}-d\boldsymbol{\Lambda})\cdot d\boldsymbol{\Lambda}_{t}\cdot(\tilde{\boldsymbol{\upmu}}_t-c\mathbf{1}_n)
\]
as $ \boldsymbol{\uppi}_{t-}(\boldsymbol{I}+d\boldsymbol{Q}_t) d\boldsymbol{\Lambda}_t = d\boldsymbol{\Psi}_t$ and 
$\boldsymbol{\uppi}_{t-}=\boldsymbol{\uppi} \cdot\prodi_{[s,t)}(\boldsymbol{I}+d\boldsymbol{Q})(\boldsymbol{I}-d\boldsymbol{\Lambda})$, and 
\[
\frac{d\boldsymbol{\uppi}}{ds}=\boldsymbol{\uppi}_{s-}\cdot \mathbf{q}_{s}-\boldsymbol{\psi}_{s}.
\]

If $s$ is an atomic point of $\Psi_{i}$, then $\pi_{i}(s-)>0$,
and letting $\Delta s\rightarrow0$ leads to the Hamilton-Jacobi-Bellman equations 
\begin{align} \label{eq: HJB point mass}
V_{s-}(\boldsymbol{\uppi}) & =  \sup_{\boldsymbol{\Lambda}(\{s\})}\left\{ V_{s}(\boldsymbol{\uppi}(I-\boldsymbol{\Lambda}(\{s\})))+\pi\boldsymbol{\Lambda}(\{s\})(\tilde{\boldsymbol{\upmu}}_s-c\cdot\mathbf{1}_n)\right\} \\
 & = \sup_{\boldsymbol{\Lambda}(\{s\})} \Big\{  \boldsymbol{\uppi}\boldsymbol{\Lambda}(\{s\})\cdot(\tilde{\boldsymbol{\upmu}}_s-c\cdot\mathbf{1}_n) \nonumber\\ 
 & +\int_{(s,T]}\boldsymbol{\uppi} \cdot \Prodi_{[s,t)}(\boldsymbol{I}+d\boldsymbol{Q})(\boldsymbol{I}-d\boldsymbol{\Lambda})\cdot d\boldsymbol{\Lambda}_{t}\cdot(\tilde{\boldsymbol{\upmu}}_t-c\cdot\mathbf{1}_n) \Big\}, \nonumber
\end{align}
where $\boldsymbol{\Lambda}(\{s\})$ satisfies
\[
0\le\boldsymbol{\uppi}\boldsymbol{\Lambda}(\{s\})\leq \boldsymbol{\uppi}.
\]

We claim that if a function $V$ satisfies \eqref{eq: HJB for V} and \eqref{eq: HJB point mass}, depending on whether $\boldsymbol{\Psi}$ is absolutely continuous or singular to the Lebesgue measure at $s$, then $V$ is the value function as defined in \eqref{eq: value fn} for all waiting time $s \in [0,T]$ and distribution $\boldsymbol{\uppi} = (\pi_1,\dots,\pi_n)$ of initial states.
To show this, we first prove $V$ is an upper bound of the value function.
Since 
\begin{multline*}
V_{(s+\Delta s)-}(\boldsymbol{\uppi}_{(s+\Delta s)-}) = V(s-, \boldsymbol{\uppi}_{s-}) \\
+ \int_{[s,{s+\Delta s}) \cap B^c} \left( \frac{\partial V}{\partial s} + \frac{\partial V}{\partial\boldsymbol{\uppi}}\cdot\frac{d\boldsymbol{\uppi}}{ds} \right)ds\\
+ \sum_{u\in [s,{s+\Delta s}) \cap B}  (V(u, \boldsymbol{\uppi}_{u}) - V(u-, \boldsymbol{\uppi}_{u-})),
\end{multline*}
with \eqref{eq: HJB for V} and \eqref{eq: HJB point mass}, the left side of the above equation is no greater than 
\[
V(s-, \boldsymbol{\uppi}_{s-}) - \int_{[s,s+\Delta s)}(d\boldsymbol{\Psi}^{ac}_{t}+d\boldsymbol{\Psi}^{s}_{t})\cdot(\tilde{\boldsymbol{\upmu}}_t-c\cdot \mathbf{1}_n)
\]
with any $\boldsymbol{\Psi}$. Letting $s+\Delta s = T$ leads to 
\begin{equation} \label{eq: verify}
V(s-, \boldsymbol{\uppi}_{s-}) \geq \int_{[s,T)}(d\boldsymbol{\Psi}^{ac}(t)+d\boldsymbol{\Psi}^{s}(t))\cdot(\tilde{\boldsymbol{\upmu}}_t-c\cdot \mathbf{1}_n), 
\end{equation}
which verifies that $V$ is an upper bound of the value function.

Now we show the equality can be achieved in \eqref{eq: verify} by providing the solution form of $\boldsymbol{\Psi}$ leading to $V$ that satisfies \eqref{eq: HJB for V} and \eqref{eq: HJB point mass}, and conclude that $V$ is indeed the value function.

From the Hamilton-Jacobi-Bellman equations \eqref{eq: HJB for V}, 
\begin{multline*}
-\frac{\partial V}{\partial s}
=\sup_{\{\boldsymbol{\Psi}_{t}:s\le t<s+\Delta s\}}\{(\tilde{\boldsymbol{\upmu}}_s-\boldsymbol{\upeta}^{c}(s+)-c\,(\mathbf{1}_n-\boldsymbol{\upgamma}^{c}(s+))\cdot\boldsymbol{\psi}_{s}\\
+(\boldsymbol{\upeta}^{c}(s+)-c \, \boldsymbol{\upgamma}^{c}(s+))\cdot\boldsymbol{\uppi}(s-)\cdot \boldsymbol{q}(s)\},
\end{multline*}
and the supremum on the right side is obtained by taking 
\begin{equation}
\psi_{i}^{c}(s)=\begin{cases}
\infty, & \text{if }\frac{\tilde{\mu}_{i}(s)-\eta_{i}^{c}(s+)}{1-\gamma_{i}^{c}(s+)}>c\text{ and }\pi_{x}(s-)>0,\\
\sum_{x\in\mathcal{X}}\pi_{x}(s-)q_{x,i}(s), & \text{if }\frac{\tilde{\mu}_{i}(s)-\eta_{i}^{c}(s+)}{1-\gamma_{i}^{c}(s+)}>c\text{ and }\pi_{x}(s-)=0,\\
0, & \text{if }\frac{\tilde{\mu}_{i}(s)-\eta_{i}^{c}(s+)}{1-\gamma_{i}^{c}(s+)}\leq c.
\end{cases}\label{eq: soln ac}
\end{equation}

Note on the right side of \eqref{eq: HJB point mass}
\begin{equation}
  \boldsymbol{\uppi}\boldsymbol{\Lambda}(\{s\})\cdot(\tilde{\boldsymbol{\upmu}}_s-c\cdot\mathbf{1}_n) \nonumber
+\int_{(s,T]}\boldsymbol{\uppi} \cdot \Prodi_{[s,t)}(\boldsymbol{I}+d\boldsymbol{Q})(\boldsymbol{I}-d\boldsymbol{\Lambda})\cdot d\boldsymbol{\Lambda}_{t}\cdot(\tilde{\boldsymbol{\upmu}}_t-c\cdot\mathbf{1}_n) 
\end{equation}
is a linear functional of $ \boldsymbol{\uppi} \boldsymbol{\Lambda}(\{s\})$ with coefficient 
\[
\tilde{\boldsymbol{\upmu}}_s-\boldsymbol{\upeta}^{c}(s+)-c\,(\mathbf{1}_n-\boldsymbol{\upgamma}^{c}(s+)),
\]
indicating the optimal $\boldsymbol{\Lambda}^{c}$ at any $s \in [0,T]$ should satisfy
\begin{equation} \label{eq: soln s}
\boldsymbol{\Lambda}^{c}(\{s\})=I(\boldsymbol{\upvarphi}^{c}(s)>c).
\end{equation}
Combining the absolutely continuous and singular solutions \eqref{eq: soln ac} and \eqref{eq: soln s} gives the desired solution form. 

Finally, we show ${\eta}_{i}^{c}(s+)$ is indeed the expected life gain after waiting time $s$ if the occupancy is concentrated in state $i$ at $s-$ and none is transplanted at $s$. Let $\boldsymbol{e}_{i}$ denote the $n$-dimensional vector with the $i$th element being $1$ and all other elements being $0$, then
\[
{\eta}_{i}^{c}(s+) =\boldsymbol{e}_{i}\cdot\int_{(s,T]}\Prodi_{(s,t)}(\boldsymbol{I}+d\boldsymbol{Q})(\boldsymbol{I}-d\boldsymbol{\Lambda}^{c})\cdot d\boldsymbol{\Lambda}^{c}_{t}\cdot\tilde{\boldsymbol{\upmu}}_t.
\]
Since for any $t\in(s,T]$,
$\boldsymbol{e}_{i}\cdot\prodi_{(s,t)}(\boldsymbol{I}+d\boldsymbol{Q})(\boldsymbol{I}-d\boldsymbol{\Lambda}^{c})$ is the limiting average occupancy at waiting time $t-$ if started from full occupancy of state $i$ at $s-$ and if none is transplanted at $s$, we have the desired result. The result for ${\gamma}_{i}^{c}(s+)$ can be shown similarly.
\end{proof}

\subsubsection*{Proof of Theorem \ref{thm: monotone}}

\begin{proof} 

By \eqref{eq: indicator phi}, it suffices to show that for
all $s\in[0,T]$ and $i\in\mathcal{X}$, 
\[
\{\tilde{\boldsymbol{\upmu}}_s-c_{1}\mathbf{1}_n\}_{i}>\{\boldsymbol{\upeta}^{c_{1}}(s+)-c_{1}\boldsymbol{\upgamma}^{c_{1}}(s+)\}_{i}
\]
is a sufficient condition for 
\[
\{\tilde{\boldsymbol{\upmu}}_s-c_{2}\mathbf{1}_n\}_{i}>\{\boldsymbol{\upeta}^{c_{2}}(s+)-c_{2}\boldsymbol{\upgamma}^{c_{2}}(s+)\}_{i},
\]
where $\{\cdot\}_{i}$ indicates the $i$th element of the vector.
Note for $j=1,2$, 
\begin{align*}
& \{\boldsymbol{\upeta}^{c_{j}}(s+)-c_{j}\boldsymbol{\upgamma}^{c_{j}}(s+)\}_{i}\\
& =\boldsymbol{e}_{i}\cdot\int_{(s,T]}\Prodi_{(s,t)}(\boldsymbol{I}+d\boldsymbol{Q})(\boldsymbol{I}-d\boldsymbol{\Lambda}^{c_{j}})\cdot d\boldsymbol{\Lambda}^{c_{j}}(t)\cdot(\tilde{\boldsymbol{\upmu}}_t-c_{j}\cdot\mathbf{1}_n),
\end{align*}
where $\{\boldsymbol{\Lambda}^{c_{j}}(t):t\in(s,T]\}$ are characterized in \eqref{eq: soln prodi cont} and \eqref{eq: soln when pi is zero}. 

A key observation is that for any $s\in[0,T]$, the form of $\{\boldsymbol{\Lambda}^{c_{j}}(t):t\in(s,T]\}$
is independent of the initial condition $\boldsymbol{\uppi}_{s}$. Therefore for any
$i\in\mathcal{X}$ and $s\in[0,T]$, $\{\boldsymbol{\Lambda}^{c_{j}}(t):t\in(s,T]\}$
also maximizes
\begin{equation*} \label{eq: obj proof mono}
\boldsymbol{e}_{i}\cdot\int_{(s,T]}\Prodi_{(s,t)}(\boldsymbol{I}+d\boldsymbol{Q})(\boldsymbol{I}-d\boldsymbol{\Lambda}^{c_{j}})\cdot d\boldsymbol{\Lambda}^{c_{j}}(t)\cdot (\tilde{\boldsymbol{\upmu}}_t - c_j \cdot\mathbf{1}_n),
\end{equation*}
which is the objective under penalty parameter $c_{j}$ if only waiting time greater than $s$ is considered and all of the occupancy is concentrated at state $i$ at time $s$,
and thus
\begin{align} \label{eq: ineq proof mono}
& \{\boldsymbol{\upeta}^{c_{1}}(s+)-c_{1}\boldsymbol{\upgamma}^{c_{1}}(s+)\}_{i}\\
&\geq \boldsymbol{e}_{i}\cdot\int_{(s,T]}\Prodi_{(s,t)}(\boldsymbol{I}+d\boldsymbol{Q})(\boldsymbol{I}-d\boldsymbol{\Lambda}^{c_{2}})\cdot d\boldsymbol{\Lambda}^{c_{2}}(t)\cdot(\tilde{\boldsymbol{\upmu}}_t-c_{1}\cdot\mathbf{1}_n). \nonumber
\end{align}
Subtracting $\{\boldsymbol{\upeta}^{c_{2}}(s+)-c_{2}\boldsymbol{\upgamma}^{c_{2}}(s+)\}_{i}$
from both sides of \eqref{eq: ineq proof mono} gives 
\begin{align} \label{eq: ineq proof mono2}
& \{\boldsymbol{\upeta}^{c_{1}}(s+)-c_{1}\boldsymbol{\upgamma}^{c_{1}}(s+)\}_{i}
-\{\boldsymbol{\upeta}^{c_{2}}(s+)-c_{2}\boldsymbol{\upgamma}^{c_{2}}(s+)\}_{i}\\
& \geq \boldsymbol{e}_{i}\cdot\int_{(s,T]}\Prodi_{(s,t)}(\boldsymbol{I}+d\boldsymbol{Q})(\boldsymbol{I}-d\boldsymbol{\Lambda}^{c_{2}})\cdot d\boldsymbol{\Lambda}^{c_{2}}(t)\cdot(c_{2}\cdot\mathbf{1}_n-c_{1}\cdot\mathbf{1}_n). \nonumber
\end{align}
The term on the right side of \eqref{eq: ineq proof mono2} is no less than $c_{2}-c_{1}$ as the probability
of future transplantation
\[
\boldsymbol{e}_{i}\cdot\int_{(s,T]}\Prodi_{(s,t)}(\boldsymbol{I}+d\boldsymbol{Q})(\boldsymbol{I}-d\boldsymbol{\Lambda}^{c_{2}})\cdot d\boldsymbol{\Lambda}^{c_{2}}(t)\cdot\mathbf{1}_n\leq1.
\]
Therefore 
\begin{equation*} \label{eq: monotone_cond}
\boldsymbol{\upeta}^{c_{1}}(s+)+c_{1}(\mathbf{1}_n-\boldsymbol{\upgamma}^{c_{1}}(s+))
\geq
\boldsymbol{\upeta}^{c_{2}}(s+)+c_{2}(\mathbf{1}_n-\boldsymbol{\upgamma}^{c_{2}}(s+)),
\end{equation*}
and the desired result follows.
\end{proof}

\subsubsection*{Proof of Theorem \ref{thm: soln_multi}}

\begin{proof}
	
	The proof is similar to that of Theorem \ref{thm: soln}. For the multi-organ-type objective \eqref{eq: multi organ obj} define the value function 
	\begin{equation*} \label{eq: mult value fn}
	V_{s-}(\boldsymbol{\uppi})
	=\sup_{\{\boldsymbol{\Psi}^{(w)}_{t}:s\leq t\leq T, \, w = 1, \dots, N_o\}} 
	\Big\{\sum_{w=1}^{N_o} \int_{[s,T]} d\boldsymbol{\Psi}^{(w)}_{t}\cdot(\tilde{\boldsymbol{\upmu}}^{(w)}_t-c^{(w)} \cdot\mathbf{1}_n)\Big\}.
	\end{equation*}	
	The corresponding Hamilton-Jacobi-Bellman equations for the absolutely continuous case is
	\begin{multline} \label{eq: mult bellman ac}
	-\frac{\partial V}{\partial s} 
	=\sup_{\{\boldsymbol{\Psi}^{(w)}_{t}:s\le t<s+\Delta s, \, w = 1,\dots, N_o\}}
	\big\{\sum_{w}\boldsymbol{\psi}^{(w)}_{s}
	(\tilde{\boldsymbol{\upmu}}^{(w)}_s-c^{(w)} \mathbf{1}_n)
	+\frac{\partial V}{\partial\boldsymbol{\uppi}}\cdot\frac{d\boldsymbol{\uppi}}{ds}\big\},
	\end{multline}
	where
	\[
	\frac{\partial V}{\partial\boldsymbol{\uppi}}
	=\int_{(s,T]}\Prodi_{(s,t)}(\boldsymbol{I}+d\boldsymbol{Q})(\boldsymbol{I}-\sum_w d\boldsymbol{\Lambda}^{(w)})\cdot \sum_w [d\boldsymbol{\Lambda}^{(w)}_{t}
	(\tilde{\boldsymbol{\upmu}}^{(w)}_t-c^{(w)} \mathbf{1}_n)]
	\]
	and
	\[
	\frac{d\boldsymbol{\uppi}}{ds}=\boldsymbol{\uppi}_{s-}\cdot \mathbf{q}_{s}-\sum_w \boldsymbol{\psi}^{(w)}_{s},
	\]
	and for the singular case is  
	\begin{align} \label{eq: mult bellman s}
	V_{s-}(\boldsymbol{\uppi}) 
	& = \sup_{\boldsymbol{\Lambda}^{(w)}(\{s\})} \Big\{  \boldsymbol{\uppi}\boldsymbol{\Lambda}^{(w)}(\{s\})\cdot(\tilde{\boldsymbol{\upmu}}^{(w)}_s-c^{(w)} \mathbf{1}_n) \\ 
	& +\int_{(s,T]}\boldsymbol{\uppi} \cdot \Prodi_{[s,t)}(\boldsymbol{I}+d\boldsymbol{Q})(\boldsymbol{I}-\sum_w d\boldsymbol{\Lambda}^{(w)})\cdot 
	\sum_w [d\boldsymbol{\Lambda}^{(w)}_{t}(\tilde{\boldsymbol{\upmu}}^{(w)}_t-c^{(w)} \mathbf{1}_n)] \Big\}. \nonumber
	\end{align}
	The forms of the optimal  $\{ \boldsymbol{\psi}^{(w)}_{s}: w = 1, \dots, N_o\}$  and $\{\boldsymbol{\Psi}^{(w)}(\{s\}) : w = 1, \dots, N_o \}$ can be readily derived from the HJB equations and the constraints \eqref{eq: mult organ constraint 1} and \eqref{eq: mult organ constraint 2}.
\end{proof}

\subsubsection*{Proof of Propositions \ref{thm: weak monotone 1} and \ref{thm: weak monotone 2}}
It suffices to show that if
\begin{align}
\tilde{\upmu}^{(w)}_i(s) - c^{(w)}_1 & > \sum_{w=1}^{N_o}  
\big[ \eta_{i}^{(w),c_1}(s+) - c^{(w)}_1 \gamma_{i}^{(w),c_1}(s+) \big] , \label{ineq: multi mono 1a}\\
\tilde{\upmu}^{(w)}_i(s) - c^{(w)}_1  & > \tilde{\upmu}^{(v)}_i(s) - c^{(v)}_1, \, \forall v \neq w, \label{ineq: multi mono 1b}
\end{align}
then
\begin{align}
\tilde{\upmu}^{(w)}_i(s) - c^{(w)}_2 & > \sum_{w=1}^{N_o}  
\big[ \eta_{i}^{(w),c_2}(s+) - c^{(w)}_2 \gamma_{i}^{(w),c_2}(s+) \big] , \label{ineq: multi mono 2a}\\
\tilde{\upmu}^{(w)}_i(s) - c^{(w)}_2 & > \tilde{\upmu}^{(v)}_i(s) - c^{(v)}_2, \, \forall v \neq w, \label{ineq: multi mono 2b}
\end{align}
where $c^{(w)}_1 - c^{(w)}_2 = \text{const} > 0$ for all $w$ (Propositions \ref{thm: weak monotone 1}), or
$c^{(w)}_1 > c^{(w)}_2$ and $c^{(v)}_1 = c^{(v)}_2$ for all $v \neq w$ (Propositions \ref{thm: weak monotone 2}). 

It is obvious that \eqref{ineq: multi mono 1b} implies \eqref{ineq: multi mono 2b} with either condition on $\{c^{(w)}\}$.
Similar to the proof of Theorem \ref{thm: monotone}, we have
\begin{multline} \label{eq: proof mult monotone}
\sum_{w=1}^{N_o}  
\big[ \eta_{i}^{(w),c_1}(s+) - c^{(w)}_1 \gamma_{i}^{(w),c_1}(s+) \big] \\
\geq
\boldsymbol{e}_{i}\cdot\int_{(s,T]}\Prodi_{(s,t)}(\boldsymbol{I}+d\boldsymbol{Q})(\boldsymbol{I}-\sum_w d\boldsymbol{\Lambda}^{(w),c_{2}})
\cdot \sum_w d\boldsymbol{\Lambda}^{(w),c_{2}}(t)
\cdot(\tilde{\boldsymbol{\upmu}}^{(w)}_t - c^{(w)}_{1}\cdot\mathbf{1}_n).
\end{multline}
Subtracting 
$\sum_{w=1}^{N_o}  
\big[ \eta_{i}^{(w),c_2}(s+) - c^{(w)}_2 \gamma_{i}^{(w),c_2}(s+) \big]$
from both sides of \eqref{eq: proof mult monotone} leads to the desired result, as
the right side of the inequality becomes 
\begin{equation*}
(c^{(w)}_{2} - c^{(w)}_{1}) \cdot \boldsymbol{e}_{i}\int_{(s,T]}\Prodi_{(s,t)}(\boldsymbol{I}+d\boldsymbol{Q})(\boldsymbol{I}-\sum_w d\boldsymbol{\Lambda}^{(w),c_{2}})
\cdot \sum_w d\boldsymbol{\Lambda}^{(w),c_{2}}(t)
\cdot \mathbf{1}_n
\geq c^{(w)}_{2} - c^{(w)}_{1}.
\end{equation*}

\bibliographystyle{imsart-nameyear}
\bibliography{ref_2019Dec}

\end{document}